\pdfoutput=1

\documentclass[final, 1p, times]{elsarticle}

\usepackage{amssymb}

\usepackage[utf8]{inputenc}
\usepackage{hyperref, natbib, amsmath, graphicx}
\usepackage[
type={CC},
modifier={by-nc-nd},
version={4.0},
]{doclicense}
{

\journal{Applied Energy}

\begin{document}

\begin{frontmatter}



\title{An optimal home energy management system for modulating heat pumps and photovoltaic systems}


\author[POM]{Lissy Langer\corref{cor1}}
\author[POM]{Thomas Volling}

\address[POM]{Work Group Production and Operations Management (POM), Technische Universität Berlin, Straße des 17. Juni 135, 10623 Berlin, Germany}
\cortext[cor1]{langer@pom.tu-berlin.de}

\begin{abstract}
Efficient residential sector coupling plays a key role in supporting the energy transition. In this study, we analyze the structural properties associated with the optimal control of a home energy management system and the effects of common technological configurations and objectives. We conduct this study by modeling a representative building with a modulating air-sourced heat pump, a photovoltaic~(PV) system, a battery, and thermal storage systems for floor heating and hot-water supply. In addition, we allow grid feed-in by assuming fixed feed-in tariffs and consider user comfort. In our numerical analysis, we find that the battery, naturally, is the essential building block for improving self-sufficiency. However, in order to use the PV surplus efficiently grid feed-in is necessary. The commonly considered objective of maximizing self-consumption is not economically viable under the given tariff structure; however, close-to-optimal performance and significant reduction in solution times can be achieved by maximizing self-sufficiency. Based on optimal control and considering seasonal effects, the dominant order of PV distribution and the target states of charge of the storage systems can be derived. Using a rolling horizon approach, the solution time can be reduced to less than 1\,min (achieving a time resolution of 1\,h per year). By evaluating the value of information, we find that the common value of 24\,h for the prediction and control horizons results in unintended but avoidable end-of-horizon effects. Our input data and mixed-integer linear model developed using the Julia JuMP programming language are available in an open-source manner.
	
\end{abstract}

\doclicenseThis




\begin{keyword}
Heat pump \sep Photovoltaics~(PV) \sep Demand-side flexibility \sep Thermal energy storage \sep Model predictive control~(MPC) \sep Mixed-integer linear programming~(MILP)
\end{keyword}

\end{frontmatter}


\section{Introduction} \label{sec:Introduction}
The German federal government’s \textit{Climate Action Plan 2050} describes the essential steps required for decarbonizing the energy and building sector in Germany~\cite{BMUB2016}. The target for the building sector is to develop a virtually climate-neutral housing stock by 2050. Because of the longevity of buildings, emission reductions of approximately 66\% are necessary by 2030 when compared with the emissions in 1990. Hence, from 2020, low-interest loans and investment subsidies of up to 45\% will be offered to the homeowners investing in renewable heating~\cite{BMWI2019}.

Thus, the goal is to improve the energy efficiency and the share of renewable energy sources with respect to the total energy consumption~\cite{BMUB2016}. In 2019, the photovoltaic (PV) capacity in Germany increased by 8\% and became almost 50\,GWp~\cite{BMWI2019a}. Approximately 76\% of the PV systems are installed in the residential sector and are smaller than 10 kWp~\cite{Bundesnetzagentur2020a}. Most of the energy consumption in the residential sector can be attributed to the thermal demand and not electricity demand (69\% for heating and 15\% for hot-water supply)~\cite{StatistischesBundesamt2019}. Therefore, efficient conversion technologies that exploit the full potential of residential PV systems are needed~\cite{Nolting2019}. Heat pumps, especially when combined with thermal storage systems, are considered to be key facilitators because they provide an efficient power-to-heat ratio [12, 17]. In 2018, 44\% of all the newly approved residential buildings in Germany were equipped with a heat pump (82\% of which were air-sourced)~\cite{StatistischesBundesamt2019a, Bdh2019}. Öko-Institut and Agora Energiewende predict that 3--5 million heat pumps will be installed in Germany alone by 2030~\cite{Agora2017, oekoinstitut2019}.

A major challenge associated with the usage of renewable energy sources is their highly uncertain and intermittent nature. The fluctuations in supply can be mitigated by demand-side flexibility; this is typically achieved in the residential sector by coupling electricity and heat~\cite{Nolting2019,Pean2019,Salpakari2016}. Thus, the thermal and battery storage systems are combined with heat pumps operating in a flexible manner with varying intensities (modulating). In Germany, every second PV system with less than 30\,kWp is already installed in combination with a similar-sized battery system~\cite{Speichermonitoring2018}.

This results in a complex control problem from an operational viewpoint. Because of the fluctuating nature of the supply of electricity from PV systems, flexibility has to be exploited in such a way that specific requirements associated with electrical energy, heating energy, and hot-water supply are met while ensuring efficient operation of the overall system. Efficient operation corresponds to profit maximization because electricity can be purchased from the grid or sold to the grid. The control problem must be solved in an integrated manner to accommodate the complex dependencies and technical constraints associated with the energy system.

The objective of this study is to better understand the optimal operation of the integrated home energy management system. We consider a single representative residential building equipped with a PV system, a modulating heat pump, thermal energy storage systems for floor heating and hot-water supply, and a battery. Based on this building, we analyze the influence of common technological configurations and objective functions on the profit- and energy-related key performance indicators~(KPIs). Thus, we identify the structural properties associated with the optimal operating strategy over a period of one year and investigate the manner in which the performance of the overall system is affected by the data forecast horizons. Therefore, we formulate a mixed-integer linear program and analyze the optimal energy flows with a time resolution of 1\,h. We model all the four system components and their interactions and particularly focus on the modulating air-source heat pump with distinct modes for floor heating and hot-water supply. In addition, we provide open-source access to our input data, the modeling parameters, the model, and the visualization code for comprehensibility and reproducibility.

The remainder of this paper is structured as follows. Section~\ref{sec:relatedwork} relates the given control problem to the literature. Section~\ref{sec:Methodology} introduces the model, and Section~\ref{sec:data} describes the model inputs in detail. Section~\ref{sec:Results} presents the results of our analysis. Section~\ref{sec:Conclusions} concludes the study and provides future research directions. \ref{sec:Parameters} presents the complete set of technical specifications and the variable and parameter nomenclature, whereas \ref{sec:Model} presents the model formulation.

\section{Related Work} \label{sec:relatedwork}

Several studies have investigated the design and control of energy systems under various objectives. Renaldi et al. (2017)~\cite{Renaldi2017} classified the previously conducted studies into two categories: (1)~studies conducted using a specific energy simulation software and (2)~studies conducted using mathematical programming methods. These studies aim to optimize the system’s design parameters or operational control under varying objectives. This study will focus on optimizing and subsequently analyzing the operational control of the underlying energy system.

The commonly used domain-specific energy simulation software packages, such as TRNSYS, ESP-r, Modelica, EnergyPlus, and IDA ICE, provide a detailed physical representation of the modeled system and can incorporate nonlinear behavior and stochastic information. However, they are computationally expensive and do not endogenously optimize the system. In addition, the simulations are often difficult to establish. Furthermore, the solutions obtained using such software are not generic but customized to the underlying energy system.

In contrast, mathematical programming methods use simplified models to generate a more generic---but less physically accurate---solution using reduced-complexity models. The formulation of heat pump models has been studied by Bloess et al. (2018)~\cite{Bloess2018}. These models can be used for optimizing the design parameters of the system or operational control. Further, a complex problem can be converted into a computationally tractable problem via simplifications. For example, the underlying demand profile is mostly modeled using external tools or simple correlations~\cite{Renaldi2017}. The accuracy and computational effort must be appropriately balanced in case of an effective model~\cite{Pean2019, Fischer2017a}. The appropriate complexity levels and model formulations of heat pump systems in different application contexts are presented by Clauss and Georges (2019)~\cite{Clauss2019}. Subsequently, we will focus on the mathematical programming approaches.

\begin{figure}[ht]
	\begin{center}
		\includegraphics[width=12 cm]{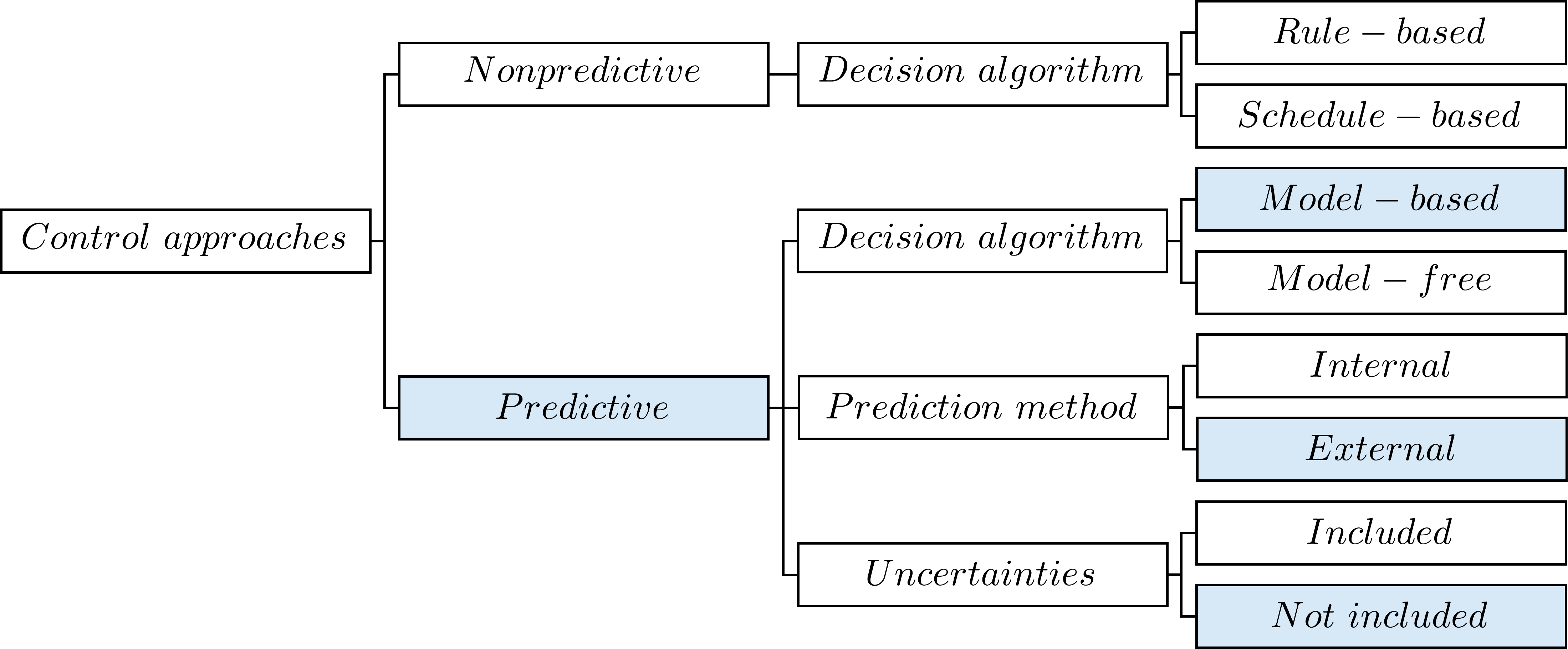}
		\caption{Classification of the control approaches for heat pumps (adapted from Fischer and Madani (2017)~\cite{Fischer2017a}). The light blue boxes indicate the classification hierarchy of this study.}
		\label{fig:SHEMS_Control_Strategies}
	\end{center}	
\end{figure}

Herein, we aim to analyze the operational control of a heat pump and PV system. Figure~\ref{fig:SHEMS_Control_Strategies} shows the comprehensive taxonomy of this problem suggested by Fischer and Madani (2017)~\cite{Fischer2017a}. They distinguished between predictive and nonpredictive approaches. The nonpredictive approaches include rule-based and schedule-based algorithms, which are based on explicit decision rules. Thus, the heat pump is activated when the room temperature, time, or some other system state reaches the specified value.

In contrast, predictive control approaches determine the operation of a heat pump based on the forecasted demand, renewable energy generation, and prices. Further, predictions can be generated internally based on past observations, or using external information sources. Uncertainties can be introduced into the analysis by varying the accuracy degree of the forecast. However, most studies assume perfect forecasts with no mismatch between the values known to the optimization model and those observed by the controlled system~\cite{Fischer2017a}. In such cases, optimal control is evaluated using the optimization model, and real-world applicability may only be partially guaranteed. It is still unclear whether this gap can be closed by more realistic (simulation) models or advanced methods of transfer learning, which transfer knowledge from one (potentially simplified) model to another model or the real world.

The predictive control approaches can be further differentiated into model-based and model-free decision algorithms. The model-free decision algorithms consider the forecasts and implement heuristics based on expert knowledge or learn by interacting with the environment (e.g., reinforcement learning systems). The model-based decision algorithms, commonly known as model predictive control (MPC) algorithms, formulate a physical world model and solve it using an exact or approximate mathematical optimization approach. A high-performing predictive approach potentially requires a large implementation effort~\cite{Fischer2017a}. In this study, we propose a predictive model-based decision algorithm by assuming that perfect information can be obtained from an external source.

Heat pump control can be categorized based on its modeling properties and targeted application (grid-based, price-based, or renewable-energy-based)~\cite{Fischer2017a}. This study belongs to the renewable-energy-based category because it predominantly analyzes the impact of integrating a heat pump with a PV system. Additional objectives include offering services to the grid or benefiting from dynamic pricing schemes. Further, we summarize the studies on integrated home energy management systems, which are most closely related to our study.

Salpakari and Lund (2016)~\cite{Salpakari2016} proposed a deterministic dynamic programming algorithm for the energy management of a low-energy house in Finland. This house was equipped with a ground-source heat pump, a PV system, a thermal storage system, batteries, and shiftable loads. They compared the performances of a rule-based self-consumption-maximizing algorithm and cost-optimal control with and without grid feed-in. They analyzed one years’ worth of data using the rolling horizon approach with a prediction and control horizon of 24\,h and a time resolution of 1\,h. They observed that the shiftable loads offer less flexibility than the thermal and battery storage systems.

Vrettos et al. (2013)~\cite{Vrettos2013} proposed a deterministic MPC algorithm for the energy management of a residential building containing a PV system, a battery, and an air-source heat pump. The demand for hot water was fulfilled using an additional electric heater. Their study focused on the cost reduction potentials based on demand-side flexibility and dynamic prices. By using the load data observed from an exemplary week in April, they observed that the energy management systems operated using the day-ahead or online price information can be beneficial for the grid and provide the owner with considerable profits. They formulated a quadratic problem and solved it using the rolling horizon approach with a prediction and control horizon of 16\,h and a time resolution of 1\,h. They assess the annual electricity costs and grid dependency for different scenarios and compare the results with those obtained from a rule-based benchmark. The grid feed-in was not considered in this case.

Fischer et al. (2017)~\cite{Fischer2017} formulated a quadratic problem for the energy management in case of a multi-family building. Their system included a PV system and an air-source heat pump with thermal storage units. They evaluated different rule-based and MPC algorithms in a dynamic simulation with an accurate heat pump model but did not provide the prediction and control horizons of the model. They modeled a set of virtual heat pumps and applied postprocessing to account for the nonlinear coefficients of performance, the separate operation modes of floor heating and hot-water supply, and the minimum compressor speeds. Further, they identified the trade-offs between thermal losses, the operational efficiency of the heat pump, and the operating costs. Their proposed MPC consistently outperformed the rule-based algorithms, although the gap could be significantly reduced by ensuring thorough calibration of the rules. In addition, their results were insensitive to the forecasting errors. Today’s values could be reasonably estimated based on yesterday’s values because of the high correlations between consecutive days. Their analysis did not include a battery and grid feed-in, and the model and simulation runtimes were not presented.

Our approach adopts and extends the heat pump model of Dengiz et al. (2019)~\cite{Dengiz2019}. They proposed two models, among which one minimizes the heating expenses without considering the electricity demand and PV generation and the other maximizes self-consumption while considering PV generation. However, both of these models did not incorporate grid feed-in or battery storage. They designed rule-based heuristics for a home energy management system with data privacy. Using scaled-up versions of the aforementioned models, they conducted a numerical study with respect to a residential area containing 40~buildings. They employed the rolling horizon regime with a prediction and control horizon of 24\,h and a time resolution of 5\,min. The KPIs of different algorithms for the two objectives are compared over 12~separate weeks during the heating period. The structure of the optimal energy flows and seasonal patterns were not presented.

After reviewing the previous studies, we identified the requirement for in-depth analysis and discussion of the optimal control of an integrated home energy management system. Until now, the impact and calibration of the essential model elements (the objective function and components, including the grid feed-in and the prediction and control horizons) remain unclear. In addition, the model runtimes are required to balance the accuracy and computational efficiency. We hope to foster further research in the field by providing an open-access dataset and model to the community, resulting in improved control algorithms, new application scenarios, and advanced policy recommendations. These issues will be discussed in the following sections.

\section{SHEMS: Model Formulation} \label{sec:Methodology}
\begin{figure}[ht]
	\begin{center}
		\includegraphics[width=10 cm]{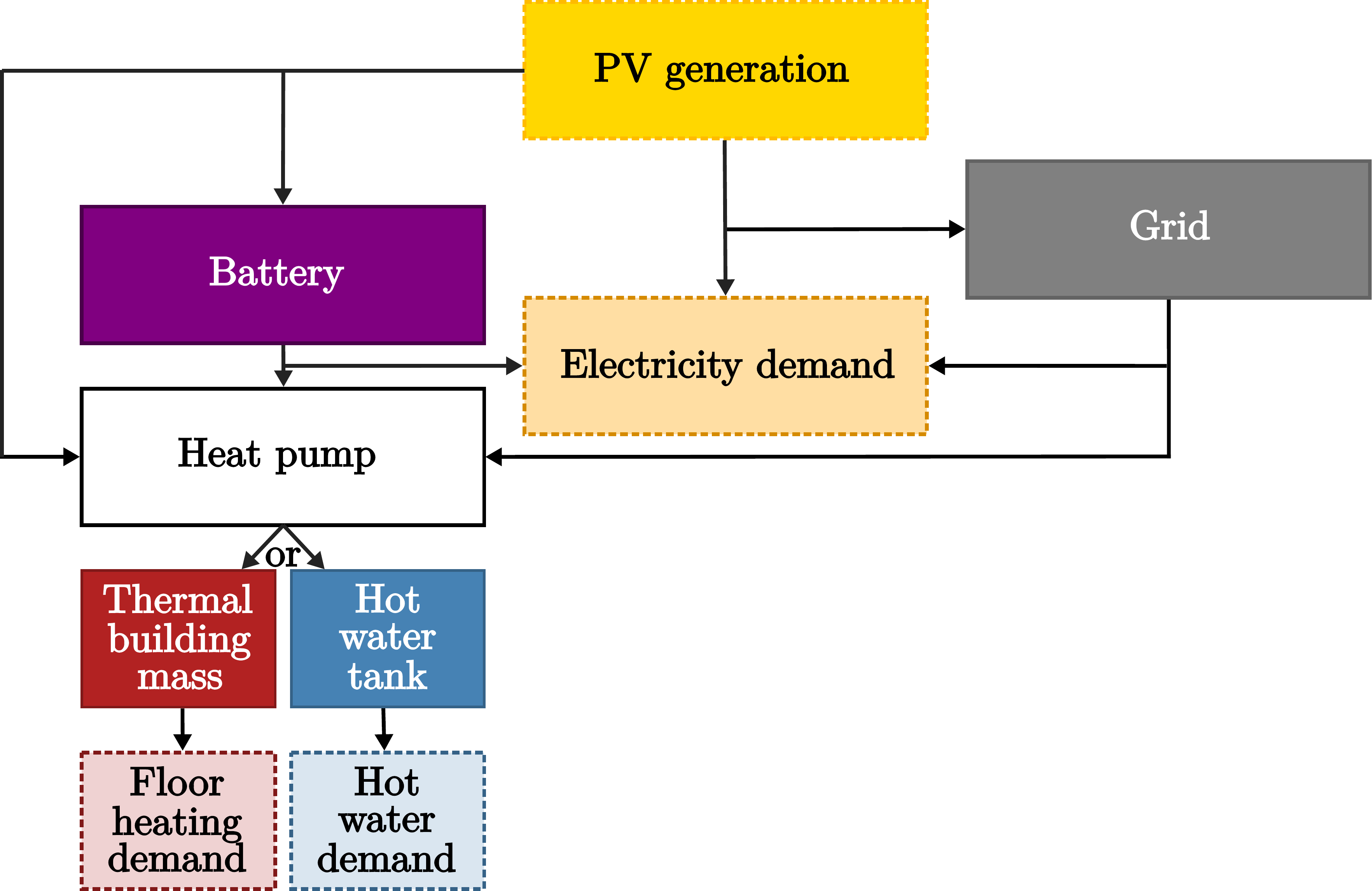}
		\caption{Energy flows of the smart home energy management system ( the parameters within the dotted frames are exogenous but known to the model).}
		\label{fig:SHEMS_Graph}
	\end{center}	
\end{figure}

In this study, we consider the smart home energy management system (SHEMS) of a single residential building (Figure~\ref{fig:SHEMS_Graph}) with a time resolution of 1\,h. The system includes a PV system~($pv$), battery~($b$), and dual-mode modulating air-source heat pump~($hp$)). At any time, the heat pump can supply heat to the floor heating system~($fh$) or the hot-water system~($hw$). The floor-heating temperature and hot-water volume must be maintained within a certain comfort range. Further, we consider two types of thermal energy buffers: the built-in thermal mass of the floor heating system and a domestic hot-water tank. The thermal buffers and battery suffer from dissipation losses. The building is connected to the power grid~($gr$), enabling the purchase of additional electricity and selling of surplus electricity.

The known parameters include the outside temperature~($t_{outside}$), PV generation~($g_e$), the demand for floor heating~($d_{fh}$), hot water~($d_{hw}$), and electricity~($d_e$), and tariffs for purchasing from the grid~($p_{buy}$) and selling to the grid~($p_{sell}$). The tariffs are assumed to be constant. In addition, the amount of exchange between the grid and building is not restricted in any direction. In this scenario, directly selling electricity from the PV system to the grid is always better than redirecting flows via battery, which results in conversion losses. Therefore, we omit the interaction between the battery and the grid. Furthermore, simultaneous charging and discharging of the battery is suboptimal because of conversion losses. Other circumstances may apply when arbitrage effects are introduced based on dynamic prices or peer-to-peer trading.

This study intends to determine the cost-optimal energy flows and heat pump operation during each hourly time period. We allow violations of the comfort ranges of the floor-heating temperature and hot-water volume but impose virtual \textit{costs} on discomfort to prevent infeasible solutions, which can occur in a monovalent system with no additional backup resistive heater. Similarly to Masy et al. (2015)~\cite{Masy2015}, we combine the two objectives using a weighted sum approach.

We formulate the problem as a mixed-integer linear program~(MILP). MILPs are popularly used because of their computational efficiency. We can holistically consider the structural properties of the optimal control policy by setting a prediction and control horizon of one year with a time resolution of 1\,h.

The system of equations is presented in \ref{sec:Model}. We describe the model components in detail in the following subsections. Section~\ref{sec:data} presents the inputs of the numerical study. The parameter settings and technical specifications are detailed in \ref{sec:Parameters}.

\subsection{Objective Function}
The net profit of the model can be maximized using the objective function of the model (Equation~\ref{eq:Objective}) by balancing the revenues obtained by selling to the grid with the cost of sourcing from the grid. In addition, there is a third term that considers the violations of the comfortable range of floor-heating temperatures and hot-water volumes. Comfort violations occur when the states of charge (SOCs) exceed the upper or lower bounds of the comfort range. Following a weighted sum approach, we add the violations weighted by the parameter $cost~factor$ (see Subsection~\ref{sec:costfactor} for details).

In our case, all positive and negative violations of the floor-heating constraint can be avoided by installing air-conditioning and a backup heater, respectively, as implemented in a previous study~\cite{Fischer2017}. Negative violations can also be resolved by increasing the capacity of the heat pump or the thermal storage system. However, designing a balanced system is beyond the scope of this study.

\subsection{Flow Constraints}
Equation~\ref{eq:DE_fulfillment} ensures that the electricity demand~($d_e$) is fulfilled at any time, either from the PV system~($X_{pv\rightarrow d_e}$), battery~($X_{b\rightarrow d_e}$), or grid~($X_{gr\rightarrow d_e}$). Meanwhile, Equation~\ref{eq:GE_restriction} ensures that the sum of flows from the PV system toward the demand~($X_{pv\rightarrow d_e}$)), battery~($X_{pv\rightarrow b}$), grid~($X_{pv\rightarrow gr}$), and heat pump~($X_{pv\rightarrow hp}$) is equal to the overall electricity generated~($g_e$). The technical specifications of the PV system are presented in Table~\ref{tab:pv_specs}.

We assume that buying from and selling to the grid is always possible. This model could also be adapted for configuring and operating an island grid; however, this adaptation is beyond the scope of our study.

\subsection{Battery Constraints}
Equation~\ref{eq:B_SOC} can be used to determine the energy balance of the battery SOC~($SOC_b$). $SOC_b$ increases with the increasing electricity supply from the PV system~($X_{pv\rightarrow b}$) after considering the conversion losses~$\eta_b$. It deceases when the battery supplies the electricity demand~($X_{b\rightarrow d_e}$) or heat pump~($X_{b\rightarrow hp}$) (with an efficiency loss of $\eta_b$). We also introduce a small dissipation loss~($loss_b$) proportional to $SOC_b$. This self-discharge is normally neglected because it is only a few percent each month~\cite{Truong2016}. However, in our case study, this self-discharge is considered for standardizing the daily load patterns. If charging at an earlier time in the day is disadvantageous, the charging time is shifted to as late as possible on the same day without significantly affecting the KPIs.

$SOC_b$ is bounded by $soc_b^{min}$ and $soc_b^{max}$, which are obtained based on the specified usable battery capacity (Equation~\ref{eq:SOC_B_min_max}). Equations~\ref{eq:B_SOC_charge_discharge} can be used to obtain the maximum charging–discharging rate~($b_{rate}^{max}$). Simultaneous charging and discharging can be avoided by the cost structure and dissipation losses. The maximum charging rates are determined based on the specified nominal power of the inverter~$b^{max}$. The technical specifications of the used battery are presented in Table~\ref{tab:battery_specs}.

\subsection{Heat Pump Constraints}\label{sec:heatpumpmodel}
To model the integrated heat pump system, we must understand its interactions with the overall home energy management system (Figure~\ref{fig:SHEMS_Graph}). The electricity for the heat pump can be supplied by the PV system~($X_{pv\rightarrow hp}$), the grid~($X_{gr\rightarrow hp}$), or the battery~($X_{b\rightarrow hp}$) (see Equation~\ref{eq:HP_balance}).

The dual operation modes are represented by separate energy flows~$X_{hp\rightarrow fh}$ and $X_{hp\rightarrow hw}$. Equations~\ref{eq:HP_mode} ensure that only one mode can be active at any given time. Therefore, we introduce the binary variable~$HP^{switch}$, which is set to one and zero when the heat pump is in the floor-heating mode and hot-water mode, respectively. Equations~\ref{eq:HP_modulation} restrict the heat pump load to its maximum power~($hp^{max}$). Here, the modulation degrees of the floor-heating and hot-water modes, denoted as $Mod_{fh}$ and $Mod_{hw}$, respectively, are continuous variables between zero and one.

In accordance with Dengiz et al. (2019)~\cite{Dengiz2019}, we model the thermal energy state of charge with respect to the mass of floor heating based on its temperature~$T_{fh}$. The floor-heating energy balance is then modeled using Equations~\ref{eq:FH_next}--~\ref{eq:FH_loss}. The next~$T_{fh}(h+1)$ is equal to the summation of the current~$T_{fh}(h)$ with the energy supplied by the heat pump~$X_{hp\rightarrow fh}$ multiplied by the coefficient of performance~$cop_{fh}$, minus the heating demand~$d_{fh}$ and the dissipation losses of the heating system~$Loss_{fh}^{+/-}$.

The coefficient of performance~$cop_{fh}(h)$ describes the ratio of the thermal power output to the electrical power input (see Subsection~\ref{sec:COP} for details). The loss term is that defined by Dengiz et al. (2019)~\cite{Dengiz2019} with slight modifications, i.e., the sign of the~$loss_{fh}$ depends on whether the outside temperature~$t_{outside}$ is greater than or lower than~$T_{fh}$. If the outdoor temperature exceeds the indoor temperature, the binary variable $Hot$ becomes one (Equations~\ref{eq:FH_hot}). In this case, the $Loss^{+/-}_{fh}$ in Equations~\ref{eq:FH_loss} becomes negative and the $loss_{fh}$ heats the system instead of cooling it. Although this binary setting simplifies the real-world situation, it prevents heating in summer through depreciation losses. All the factors influencing~$T_{fh}$ must be multiplied by a conversion factor~$conv_{fh}$ that converts kWh to $^{\circ}$C. The conversion factor, given by Equation~\ref{eq:FH_conv}, requires the hourly resolution and volume specification~$v_{fh}$ associated with floor heating (see Dengiz et al. (2019)~\cite{Dengiz2019} for details).

Equations~\ref{eq:FH_min_max} determine the comfort-range violations associated with the floor heating system. A positive comfort violation~$T_{fh}^+$ occurs when the SOC~$T_{fh}$ in a given time period exceeds the upper bound of the comfort range~$t^{max}_{fh}$. Conversely, a negative violation~$T_{fh}^-$ occurs when $T_{fh}$ is below the lower bound of the comfort range~$t^{min}_{fh}$.

The hot-water constraints (Equations~\ref{eq:HW_next}--\ref{eq:HW_min_max}) are governed by the same principles. However, the loss factor $loss_{hw}$ is a positive constant, and the conversion factor~$conv_{hw}$. converts kWh to liters. The conversion factor, given by Equation~\ref{eq:HW_conv}, requires the hourly resolution and supply temperature ~$t^{supply}_{hw}$ of the hot-water tank (see Dengiz et al. (2019)~\cite{Dengiz2019} for details). The comfort violations (Equations~\ref{eq:HW_min_max}) are given not in $^{\circ}$C but in liters because the comfort range is given as the volume of well-tempered hot water. In the objective function, these units are converted to EUR based on the cost factor.

The model formulated using Equations~\ref{eq:Objective}--~\ref{eq:HW_min_max} is an extended variant of the multiperiod capacitated transshipment problem. Furthermore, it includes binary decision variables that ensure single-mode operation of the heat pump and sign adjustment of~$Loss^{+/-}_{fh}$. Thus, the problem becomes a MILP that can be efficiently solved under the given problem sizes.

\section{SHEMS: Data Input} \label{sec:data}
We selected open-access data sources and open-source modeling frameworks to ensure the reproducibility of our model. In this section, we describe our model data and the adopted technical specifications.

\subsection{Model Implementation}
The model is written using the mathematical optimization language \href{https://github.com/JuliaOpt/JuMP.jl}{\textit{JuMP}} (package version~0.21.2) under the  \href{https://www.mozilla.org/en-US/MPL/2.0/}{\textit{MPL License}} \cite{Dunning2017}. The package is embedded in the \href{https://github.com/JuliaLang/julia}{\textit{Julia}} programming language (version~1.4.0) under the \href{https://github.com/JuliaLang/julia/blob/master/LICENSE.md}{\textit{MIT License}}. Both these software packages are open-source and available free of charge. The implementation code, data, and visualization code are available on \href{https://github.com/lilanger/SHEMS/single_building/tree/master/single_building}{\textit{GitHub}} under the \textit{MIT License}. The model is solved using the \href{https://www.gurobi.com/}{\textit{Gurobi}} solver (package version~0.7.6) under an academic license.

\subsection{Heating, Hot Water, and Electricity Demand}
To the best of our knowledge, no publicly available dataset contains the electricity demands and thermal loads associated with heating and hot water usage in single households. Accordingly, we constructed a dataset using \textit{BEopt--Building Energy Optimization with Hour-by-Hour Simulations} provided by the \textit{National Renewable Energy Laboratory}~(NREL). The software is based on \textit{EnergyPlus} simulations, and were validated by Christensen et al. (2016)~\cite{Christensen2006}. Our dataset contains the heating-demand, hot-water-demand, and electricity-demands with an aggregation level of 1\,h. We selected Chicago as the location because the NREL data include only the cities in US. Furthermore, the climate conditions in Chicago are similar to those of Northern Europe (although the solar radiation is slightly higher in Chicago in the cold seasons; compared to Berlin, autumns are milder and winters are colder). The level of temperatures at this location is suitable to evaluate the heating control---also in Northern Europe. The aggregated demand values, which are comparable to those in Germany~\cite{StatistischesBundesamt2019b}, are presented in Table~\ref{tab:demands}.

\begin{table}[ht]
	\centering
	\caption{Aggregated floor-heating, hot-water, and electricity demands in the given year.}\label{tab:demands}
	\begin{tabular}{ c c c }
		\hline
		\textbf{Floor-heating demand} & \textbf{Hot-water demand} & \textbf{Electricity demand} \\ \hline
		11880~kWh            & 3018~kWh                  & 3097~kWh                    \\
		66\%                 & 17\%                      & 17\%
	\end{tabular} 
\end{table}

We modeled a four-person three-bedroom bungalow with a total floor area of 104 m$^2$ to generate representative results. This bungalow is larger than the average house in Germany (93 m$^2$) and smaller than the houses occupied by high-income earners without children (117 m$^2$) and medium-income earners with children (122 m$^2$). These data are obtained based on the characteristic household types of the \textit{German Income and Consumption Sample}~\cite{oekoinstitut2019}.\\

\begin{figure}[ht]
	\begin{center}
		\includegraphics[width=12 cm]{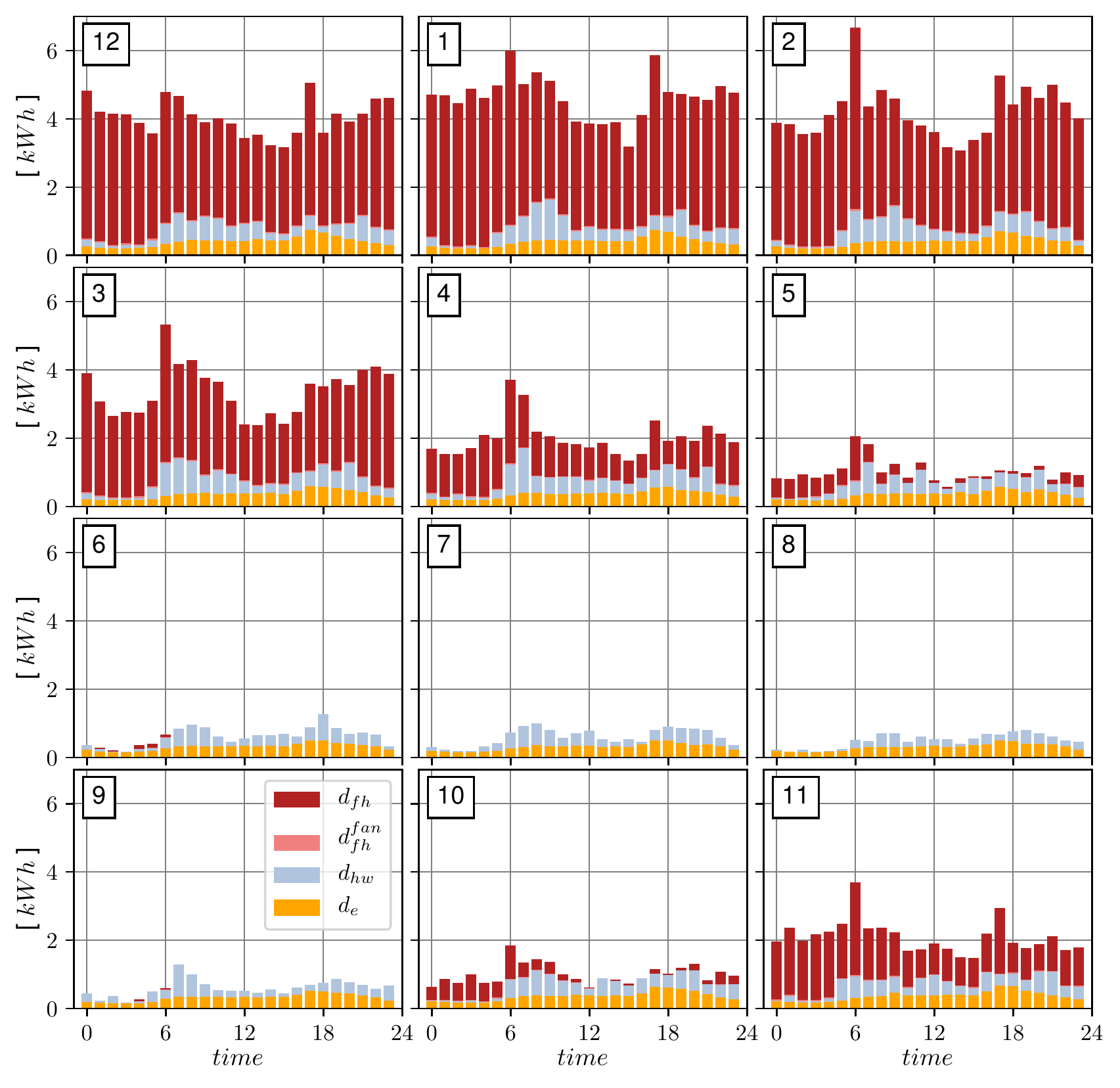}
		\caption{Electricity, floor-heating, and hot-water demand by month (stacked bars).}
		\label{fig:Heat_demand}
	\end{center}
\end{figure}

Because our objective is to control the output of a heat pump in an integrated system, thermal demands are crucial. Figure~\ref{fig:Heat_demand} shows the hourly mean values of the floor-heating and hot-water demands in each month. The floor-heating demand is very high in winter (top~row) and almost negligible in summer (third~row). The electricity and hot-water demands remain relatively constant throughout the year but follow a distinct daily pattern.

\subsection{Photovoltaic Generation and Temperature Data} \label{sec:ninja}
The photovoltaic data were extracted from \href{www.renewable.ninja}{\textit{Renewable.ninja}} licensed under \href{https://creativecommons.org/licenses/by-nc/4.0/}{\textit{CC BY-NC 4.0}}. The PV outputs were derived using the \textit{Global Solar Energy Estimator} model~\cite{Pfenninger2016}, which uses satellite observations from the \textit{NASA MERRA reanalysis}~\cite{Rienecker2011}. The temperature data were extracted from the same sources. Again, Chicago was selected as the reference location to ensure coherence with the demand profiles.

\subsection{Photovoltaic and Battery Specifications}
According to the \textit{Speichermonitoring} of \textit{RWTH Aachen}, in 2017, every second PV system smaller 30\,kWp in Germany has been installed in combination with a battery storage system~\cite{Speichermonitoring2018}. This combined system is popularly purchased for several main reasons, including hedging against the rising electricity prices, supporting the energy transition, or general interest in technology~\cite{Speichermonitoring2018}. The average capacity of the installed battery (7.8\,kWh in 2017) is trending upward, and the unit costs are decreasing. The acceptable cost has remained constant at approximately 10,000\,EUR, which is the approximate price of our Tesla Powerwall~2 with a capacity of 14\,kWh (see technical specifications in Table~\ref{tab:battery_specs}).

The ratio of battery capacity (kWh) to the installed PV capacity (kWp) was approximately 1.0 in 2017 and has increased with the increasing size of the systems. Many residential PV systems in Germany are smaller than 10\,kWp (below this limit, the renewable energy levies on self-consumption are avoided and users can profit from the high feed-in tariff)~\cite{Bundesnetzagentur2020}. As of January 2020, new installations are guaranteed a feed-in tariff of 9.87\,ct/kWh over a time horizon of 20\,years. In 2017, the mean installation size was 8.1\,kWp and was continuously increasing. We selected a 10\,kWp system because it is cost-efficient and accounts for the purchase motivations and growth trends.

\subsection{Heat Pump Specifications}
The specifications of the selected heat pump are presented in Table~\ref{tab:HP_characteristics}. This heat pump was classified using the framework presented by Fischer and Madani (2017)~\cite{Fischer2017a}. Specifically, we model a modulating (variable speed) heat pump that can be continuously adjusted (rather than a heat pump that can be merely switched on and off). The designed air-to-water heat pump has two separate storage units for floor heating and hot water. Here, the thermal building mass of the floor heating system provides the storage volume for floor heating.

The air-source heat pump was selected because it comprises approximately 82\% of all the newly installed heat pumps in Germany. This type of heating is typically combined with a floor heating system in newly constructed buildings.

Our objective is to elucidate the structure of the optimal control strategy of an integrated system. We excluded the minimum run time, part load performance, and cycling losses to simplify the model as far as possible. Although these simplifications are in agreement with the scope of this study~\cite{Madani2011}, they impose various limitations, as discussed in the concluding remarks. All the technical specifications of the heat pump are presented in Table~\ref{tab:HP_specs}.

\begin{table}[ht]
	\centering
	\caption{Heat-pump characteristics \cite{Fischer2017a}.}\label{tab:HP_characteristics}
	\begin{tabular}{p{1.2cm} p{1.2cm} p{2.2cm} p{1.4cm} p{2.7cm} p{2.2cm}}
		\hline
		\textbf{Heat}  \newline  \textbf{source}       &  \textbf{Heat} \newline \textbf{sink}  & 
		\textbf{Heat}  \newline  \textbf{distribution} &  \textbf{Heat} \newline  \textbf{storage} & 
		\textbf{Heat}  \newline  \textbf{supply concept}  & 
		\textbf{Capacity}  \newline  \textbf{control}  \\ \hline
		Air   & Water   & Underfloor & Tanks & Monovalent & Modulated
	\end{tabular} 
\end{table}

\subparagraph{Coefficient of Performance of the Heat Pump.} \label{sec:COP}
The coefficient of performance ($cop$) determines the ratio of the thermal power output to the electrical power input. The $cop$ of heat pumps is often simplified and considered to be a constant (typically, $cop=3$). Fischer at al. (2017)~\cite{Fischer2017} more accurately approximated the $cop$ by considering the temperature difference and compressor speed. They applied Taylor linearization on virtual heat pumps. In accordance with Dengiz et al.~\cite{Dengiz2019} and using manufacturers’ data, we approximate the $cop$ based on the difference between the sink and source temperatures. The evaluated parameter, given by Equation~\ref{eq:cop}, is exogenous to the model, and the model remains linear.

\begin{equation} \label{eq:cop}
	\begin{aligned}
		cop_{fh/hw}(h) = max \left\lbrace 5.8-\frac{1}{14}*|t_{fh/hw}^{supply}-t_{outside}(h)|, 0  \right\rbrace 
	\end{aligned}
\end{equation}

\begin{figure}[ht]
	\begin{center}
		\includegraphics[width=10 cm]{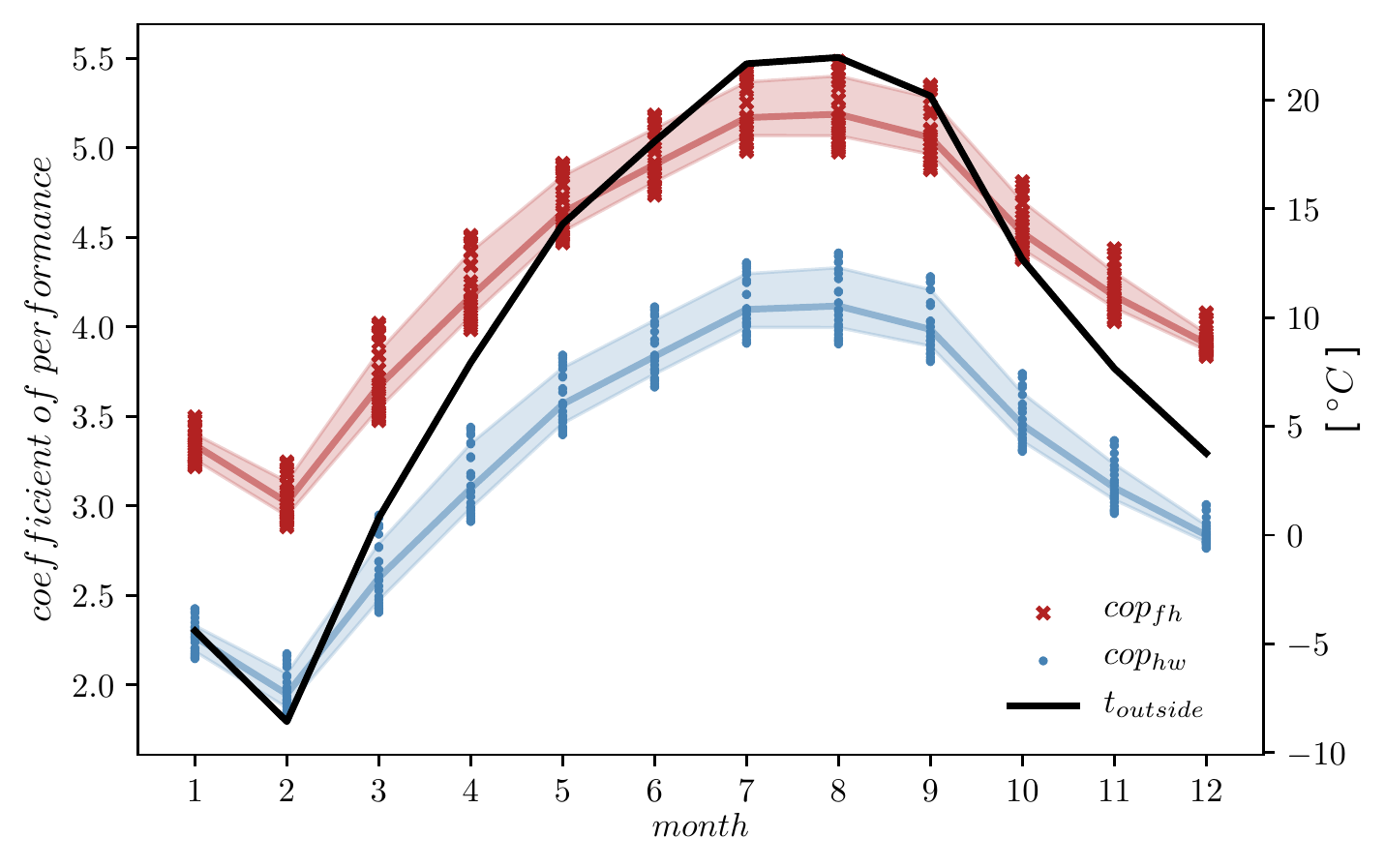}
		\caption{Coefficients of performance of the heat pump for floor heating (and hot water in a given year)(Points: hourly mean values, Lines: monthly median values, Areas: 50\% confidence interval, Black Line: mean outside temperatures).
		}
		\label{fig:COPs}
	\end{center}
\end{figure}

In accordance with Dengiz at al. (2019)~\cite{Dengiz2019}, we set the supply temperatures of the floor-heating and hot-water systems to $t_{fh}^{supply}$=\,30$^{\circ}$C and $t_{hw}^{supply}$=\,45$^{\circ}$C (using a fresh water station), respectively. As shown in Figure~\ref{fig:COPs}, $cop$ increased in summer because the gap between the supply and outside temperatures decreased. High coefficients of performance can be achieved in high-insulated buildings with floor heating, where a low temperature supply is sufficient.

\subsection{Electricity Tariffs}
Under the current regulations in Germany, most customers do not pay dynamic electricity costs or trade at the day-ahead or intraday market. Thus, our model assumes a constant PV feed-in tariff of 10\,ct/kWh and a constant electricity retail price of 30\,ct/kWh. These values are based on the current feed-in tariff and current end-consumer retail prices of green electricity~\cite{Bundesnetzagentur2020, GreenpeaceEnergy2020, NaturstromAG2020}.

In Germany, the guaranteed feed-in tariffs are continuously decreasing~\cite{Bundesnetzagentur2020}. Although the PV production costs have reduced to become approximately one-third of the retail electricity prices~\cite{Kost2018}, the guaranteed feed-in tariff barely covers the production costs. Thus, feeding one’s PV production into the grid is not profitable. Profitability can be achieved only by increasing one’s share with respect to the local energy consumption (see Subsection~\ref{sec:KPIs}).

\subsection{Cost Factor Specifications}\label{sec:costfactor}
\begin{figure}[ht]
	\begin{center}
		\includegraphics[width=10 cm]{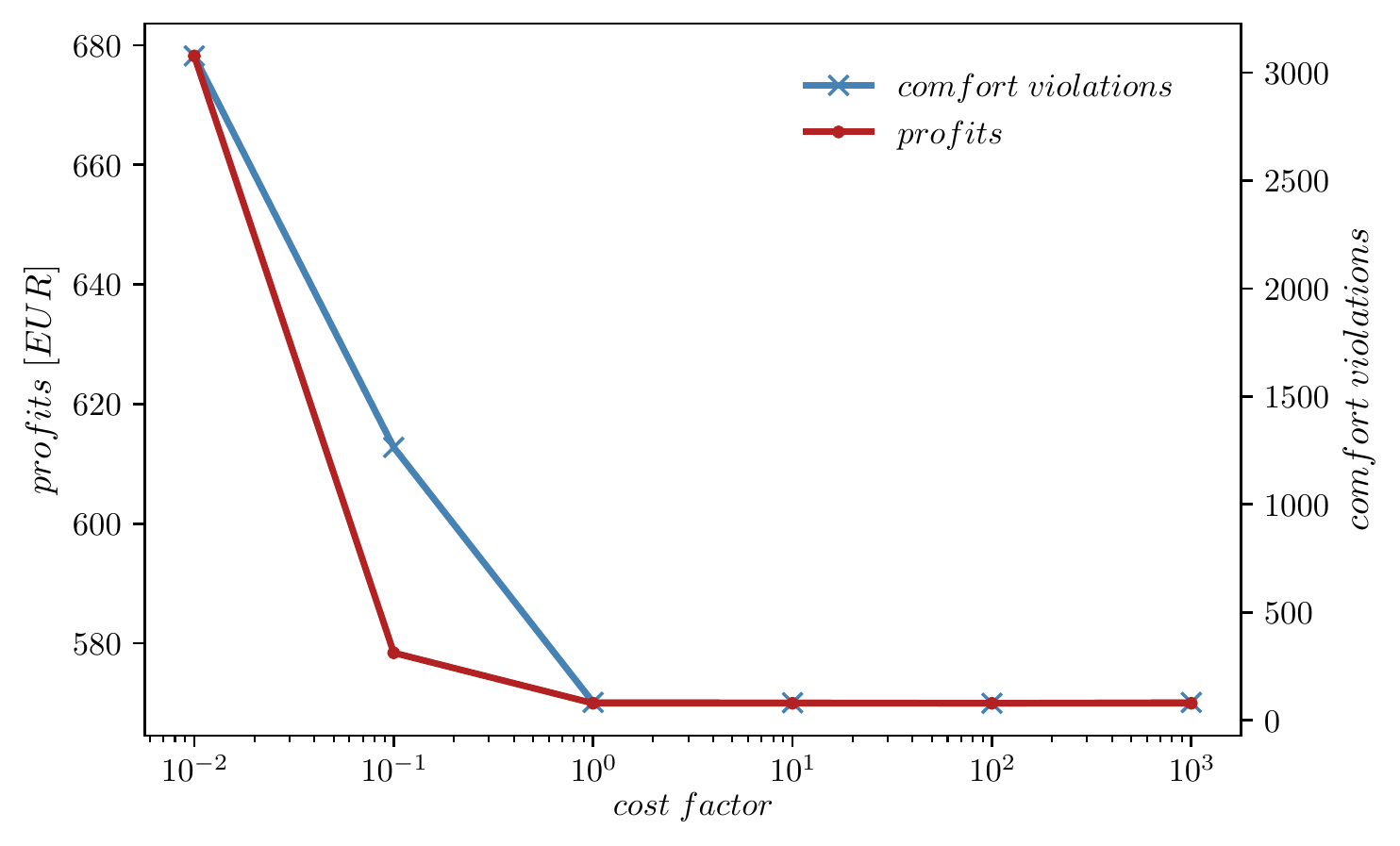}
		\caption{Numerical experiment determining the cost factor in the objective function (base case).}
		\label{fig:costfactors}
	\end{center}
\end{figure}

The cost factor parameter in the objective function weighs the importance of the comfort violations with respect to the net profits. This trade-off can be modeled using different approaches. Pean et al. (2019)~\cite{Pean2019} comprehensively reviewed various approaches. Increasing the cost factor reduces the profits and the total number of comfort violations (Figure~\ref{fig:costfactors}). Based on our preliminary experiments and a previous study~\cite{Yu2020}, we observed that a symmetric linear penalty function with a cost factor of 1 gives Pareto-optimal results.

Therefore, both the positive and negative violations of the comfort ranges with respect to floor heating and hot-water supply were equally penalized at marginal costs of 1\,EUR. Increasing the cost factor further, as in~\cite{Masy2015}, does not reduce the cumulative number of comfort violations because such comfort violations can be solely attributed to the lack of air conditioning in summer.

Based on the model in \ref{eq:Objective} to \ref{eq:HW_min_max} and the aforementioned parameter specifications, we constructed a fully specified instance of a representative residential building with the required technical equipment combination. In the following section, we will analyze the optimal behavior of the system and discuss improvements of integrated control.

\section{Results and Discussion} \label{sec:Results}
In this section, we conduct an extensive numerical analysis to understand the optimal behavior of SHEMS. Further, we investigate the potential of combining all the technological components into a single integrated control. In addition, we scrutinize the structure of the optimal control policy under seasonal effects and assess the value of information. Then, we discuss promising mechanisms for improving the rule-based control schemes.

In Subsection~\ref{sec:KPIs}1, we examine the common KPIs associated with four different technological configurations and two alternate objective functions. In the first case study, the effects of integrating a battery and the functionality of grid feed-in into the system are examined. In the second case study, two objective functions are evaluated: maximizing the self-consumption and maximizing the self-sufficiency. In Subsection~\ref{sec:Flows}, we discuss the optimal energy flows throughout the year to identify seasonal patterns in demand fulfillment under different conditions. The flow analysis is supplemented by the evaluation of the intraday effects based on illustrative edge cases during summer and winter. To design new decision rules, one must determine target or threshold values. Therefore, in Subsection~\ref{sec:Target_SoC}, we analyze the target SOCs, i.e., the levels to charge up to, of the storage systems and the schedules, i.e., the time at which charging is optimally conducted. Finally, the value of information in the rolling horizon planning approach is discussed in Subsection~\ref{sec:Rolling Horizon}. We determine the effects of adapting prediction and control time horizons, i.e., the foresight capability of the model and the number of fixed time periods in each execution step, respectively, on the overall performance of the model.

\subsection{Key Performance Indicators} \label{sec:KPIs}
The KPIs of our model were evaluated for four technological configurations, hereafter called cases, and three objective functions (Table\ref{tab:cases}).

\begin{table}[ht]
	\centering
	\caption{Configurations and objective functions of the numerical study}\label{tab:cases}
	\begin{tabular}{ p{1.8cm} p{2cm} p{8.4cm} }
		\hline
		\multicolumn{3}{l}{\textbf{Technological configurations (objective function = Profit)}}         \\ \hline
		Case 1 & Base case & Battery installed and grid feed-in allowed  \\ 
		Case 2 & No battery & No battery installed but grid feed-in allowed   \\
		Case 3 & No feed-in & Battery installed but no grid feed-in allowed  \\
		Case 4 & No both & No battery installed and no grid feed-in allowed   \\\hline
		\multicolumn{3}{l}{\textbf{Objective functions (technological configuration = Base case)}}                   \\ \hline
		Objective 1 & Profit & Minimize the net cost of grid exchange and comfort violations \\ 
		Objective 2 & Self-consumption & Minimize the grid feed-in and comfort violations (sum of violations is constrained to the base case value)  \\
		Objective 3 & Self-sufficiency & Minimize sourcing from the grid and comfort violations (sum of violations is constrained to the base case value) \\ \hline
	\end{tabular} 
\end{table}

In the base case (Case~1), the system was established based on the specifications in Section~\ref{sec:data}. The system did not contain a battery in Cases~2 and~4. In Cases~3 and~4, the feed-in tariff was zero. In all the four cases, we applied the objective function described in Section~\ref{sec:Methodology}~(Profit).

Further, we determine the effects of maximizing the self-consumption (Objective~2) and self-sufficiency (Objective~3). These objective functions are commonly used in literature~\cite{Fischer2017a}.

\begin{table}[ht]
	\centering
	\caption{Annual KPIs of the numerical study (Seco: self-consumption, Sesu: self-sufficiency).}\label{tab:KPIs_scenarios}
	\begin{tabular}{ l c| c c c|c c }
		\hline
		                                         & Case 1: & Case 2: & Case 3: & Case 4: & Obj.2 & Obj.3 \\
		Objective:                               & Profit    & Profit    & Profit    & Profit    & Seco  & Sesu  \\
		Battery:                                 & Yes     & No      & Yes     & No      &  Yes  & Yes   \\
		Grid feed-in:                            & Yes     & Yes     & No      & No      &  Yes  & Yes   \\ \hline
		Energy consumption [kWh]                 & 7515    & 7517    & 7590    & 7562    & 7701  & 7592  \\
		Self-consumption rate [\%]               & 37      & 24      & 38      & 24      &  39   & 38    \\
		Self-sufficiency rate [\%]               & 79      & 52      & 79      & 52      &  79   & 79    \\ \hline
		Overall profits [EUR]                    & 570     & 183     & -474    & -1085   &  517  & 557   \\
		PV curtailment [kWh]                     & -       & -       & 10309   & 12641   &   -   & -     \\
		Sum of violations [$^{\circ}$C] or [l] & 79      & 79      & 80      & 79      &  79   & 79    \\
		Runtime [min]                            & 47      & 39      & 14      & 17      &  36   & 10    \\ \hline
		Mean $SOC_b$ [kWh]                       & 2       & -       & 3       & -       &   7   & 4     \\
		Mean $T_{fh}$ [$^{\circ}$C]            & 21      & 21      & 21      & 21      &  21   & 21    \\
		Mean $V_{hw}$ [l]                        & 99      & 99      & 78      & 88      &  97   & 92    \\ \hline
	\end{tabular} 
\end{table}

Luthander et al. (2015)~\cite{Luthander2015} defined self-consumption as the share of locally produced electricity consumed locally. We minimize feeding into the grid to incorporate this goal in our adapted objective function (Equation~\ref{eq:obj_self-consumption}).

\begin{equation} \label{eq:obj_self-consumption}
min \sum_{h \in H} \left( X_{pv\rightarrow gr}(h) + C_{violation}(h) \right)
\end{equation} 

Self-sufficiency (or autarky) is defined as the share of local demand satisfied by local production. The reference point in this case is not the renewable generation but the local demand. Accordingly, Equation~\ref{eq:obj_self-sufficiency} minimizes sourcing from the grid.

\begin{equation}\label{eq:obj_self-sufficiency}
min \sum_{h \in H} \left( X_{gr\rightarrow d_e}(h) + X_{gr\rightarrow hp}(h) + C_{violation}(h) \right)
\end{equation}

In both the functions, we retain the original penalty terms associated with the comfort violations. However, Equation~\ref{eq:obj_constraint} constrains the sum of comfort violations with respect to the optimal value of the base case ($violations_{bc}$), determined as 79 units in our numerical experiments, to comply with the multiobjective setup and ensure comparable results.

\begin{equation}\label{eq:obj_constraint}
\sum_{h \in H} \left( T_{fh}^+(h) + T_{fh}^-(h) + V_{hw}^+(h) + V_{hw}^-(h) \right) \leq violations_{bc}
\end{equation}

We must prevent simultaneous charging and discharging of the battery in the alternate objective functions. In the base case, simultaneous charging was circumvented based on the cost structure and the dissipation losses of the battery. We replace Equation~\ref{eq:B_SOC_charge_discharge} by Equations~\ref{eq:B_SOC_switch} for analyzing the alternate objectives. The binary variable $B^{switch}(h)$ is switched on and off when the battery is charged and discharged, respectively, preventing simultaneous charging and discharging:

\begin{align}
X_{pv\rightarrow b}(h) &\leq B^{switch}(h) * b_{rate}^{max},\ \forall h 	\label{eq:B_SOC_switch} \\
X_{b\rightarrow d_e}(h) + X_{b\rightarrow hp}(h) &\leq \left(1 - B^{switch}(h)\right) * b_{rate}^{max},\ \forall h \nonumber
\end{align}

The annual aggregated results are summarized in Table~\ref{tab:KPIs_scenarios}. The overall annual energy consumption includes the electricity usage that fulfills the demand de and charges the thermal storage systems. We also report the self-consumption and self-sufficiency rates.

The overall profits refer to the net sum of the revenues and costs of grid exchanges. We exclude the virtual costs associated with comfort violations to improve the interpretability. In Cases~3 and~4, where the grid feed-in tariff was zero, PV generation could not always be processed by the system; in particular, the storage systems could not be utilized when the supply exceeded the demand. This is referred to as \textit{PV curtailment}. Further, we report the sum of the violations of the upper and lower bounds with respect to the comfort ranges for $hw$ and $fh$. Finally, we present the runtimes and average SOCs of the storage systems.

\subparagraph{Battery and Grid Feed-in.} 
Massive PV curtailment could be observed in the system in the absence of grid feed-in (Cases~3 and~4). The system could not locally utilize the PV surplus despite the optimized control and thermal storage. In Case~3, this effect was partly mitigated by the battery; therefore, the self-consumption, self-sufficiency, and profit were higher than those in Case~4. The energy consumption was slightly improved ($<$1\%). The system uses the locally generated energy instead of purchasing from the grid, increasing the dissipation losses.

In Case~2, PV curtailment was eliminated by allowing grid feed-in. This revenue stream resulted in positive net profits. Further, the energy consumption slightly decreased because of less stored energy, thereby avoiding dissipation losses. However, without a battery, grid feed-in alone exerts neither a positive nor negative effect on the self-consumption and self-sufficiency rates.

In addition, feed-in did not significantly affect the levels of comfort violations. Inherent violations of the temperature comfort range were observed during the summer because of no air conditioning.

The grid feed-in significantly changed the system behavior by monetizing the PV surplus. Therefore, it should be incorporated into the integrated energy management system if available in the considered market. Alternate distribution channels, such as peer-to-peer networks, should be installed in markets with no feed-in tariff or when tariffs are decreasing. Thus, the PV battery systems of a given size can be effectively utilized. The battery increases the profits by increasing the self-consumption and (especially) self-sufficiency rates.

The overall system performance is considerably improved using the integrated approach. However, the runtime of the model increases when considering the feed-in tariffs. In the base case, the runtime required for optimization throughout the year was approximately 50 min, almost all of which was consumed by the solver. The runtime may be significantly reduced by incorporating the initial solutions during subsequent runs. In Subsection~\ref{sec:Rolling Horizon}, the solution time can be reduced using the rolling horizon approach.

\subparagraph{Self-Consumption and Self-Sufficiency.} \label{sec:selfconsumption}
The optimization of self-consumption (Objective~2) slightly increased the self-consumption (+2\%) and reduced the runtime (-24\%). However, after the optimization of self-consumption, the energy consumption increased by 2\% and the profit decreased by 9\% when compared with those observed in the base case. The battery usage increased, as indicated by the mean battery SOC (52\%); consequently, more dissipation losses were induced. In contrast to prior studies, for example, Fischer et al. (2017)~\cite{Fischer2017}, the usage of the thermal storage systems did not change when compared with that in the base case. To ensure comparable results, the number of comfort violations was capped at that observed in the base case.

In case of home energy management systems containing a PV system, a modulating heat pump, and thermal and battery storage systems, the maximization of self-consumption introduced an unintended system behavior. This result contradicts the view associated with renewable-energy studies, such as that conducted by Fischer and Madani (2017)~\cite{Fischer2017a}, based on which self-consumption should be maximized in the objective function.

The self-sufficiency objective (Objective~3) achieved better results than Objective~2. However, the self-sufficiency rate did not increase, whereas the energy consumption increased by 1\% and the profits decreased by 2\% compared with those observed in the base case. Local PV generation is preferred over sourcing from the grid, which incurs depreciation losses. In the base case, the efficiency could be increased by selling to the grid now (to avoid storage losses) and sourcing from the grid later. The runtime with self-sufficiency optimization was 79\% lower than that in the base case and was even lower than those associated with the reduced-technical setups. 

Thus, the maximization of self-consumption results in an unintended system behavior. The self-consumption rates only slightly improved, whereas most of the other KPIs deteriorated. Maximizing the self-sufficiency did not significantly change the base case results except for the runtime, which was considerably reduced. Meanwhile, our initial cost-minimizing objective function improved the efficiency which results in reduced energy consumption and increased profits.

\subsection{Analyzing the Optimal Flows} \label{sec:Flows}
In this subsection, we analyze the optimal flows in the base case (Case~1; with a battery and grid in-feed) combined with our initial cost-minimizing objective function. This analysis will enhance our understanding of the optimal behavior of the integrated system.

\begin{figure}[ht]
	\begin{center}
		\includegraphics[width=13 cm]{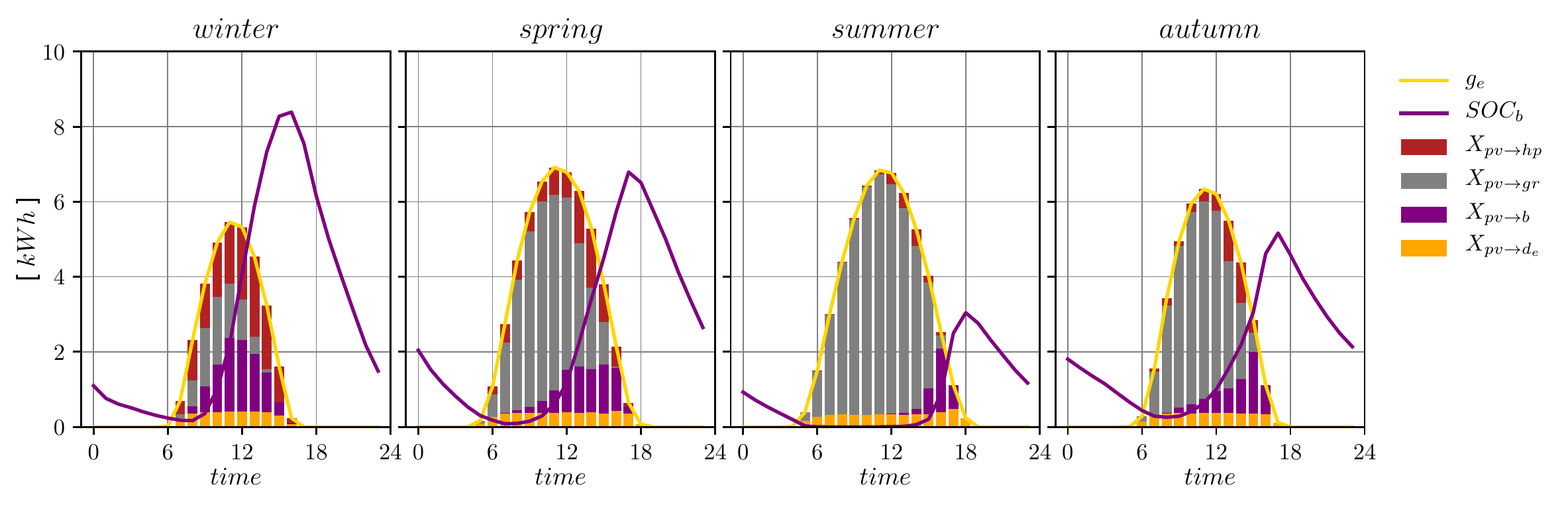}
		\caption{Mean flows from the PV system (stacked bars) and battery SOCs (purple lines) during each season (base case).}
		\label{fig:PV_Flows}
	\end{center}
\end{figure}

\subparagraph{Seasonal Flows.}
Figure~\ref{fig:PV_Flows} illustrates the seasonal mean PV generation (yellow line), the corresponding sinks (stacked bars), and the mean battery SOC (purple line). During summer, the low heating demand and high PV generation naturally increased the grid feed-in (gray bars in Figure~\ref{fig:PV_Flows}). The lower battery SOC in summer when compared with the remaining seasons will be analyzed in Subsection~\ref{sec:Target_SoC}. Generally, the battery charging was delayed toward the end of the day because of self-discharging losses. The exceptions to this rule indicate that both seasonality and time of day (both how much and when to charge) are important in rule-based approaches. However, Figure~\ref{fig:PV_Flows} shows only aggregated seasonal flows and does not allow the analysis of the daily operational patterns. In the next two paragraphs, we analyze the operational patterns of some edge cases observed during winter and summer.

\begin{figure}[ht]
	\begin{center}
		\includegraphics[width=12 cm]{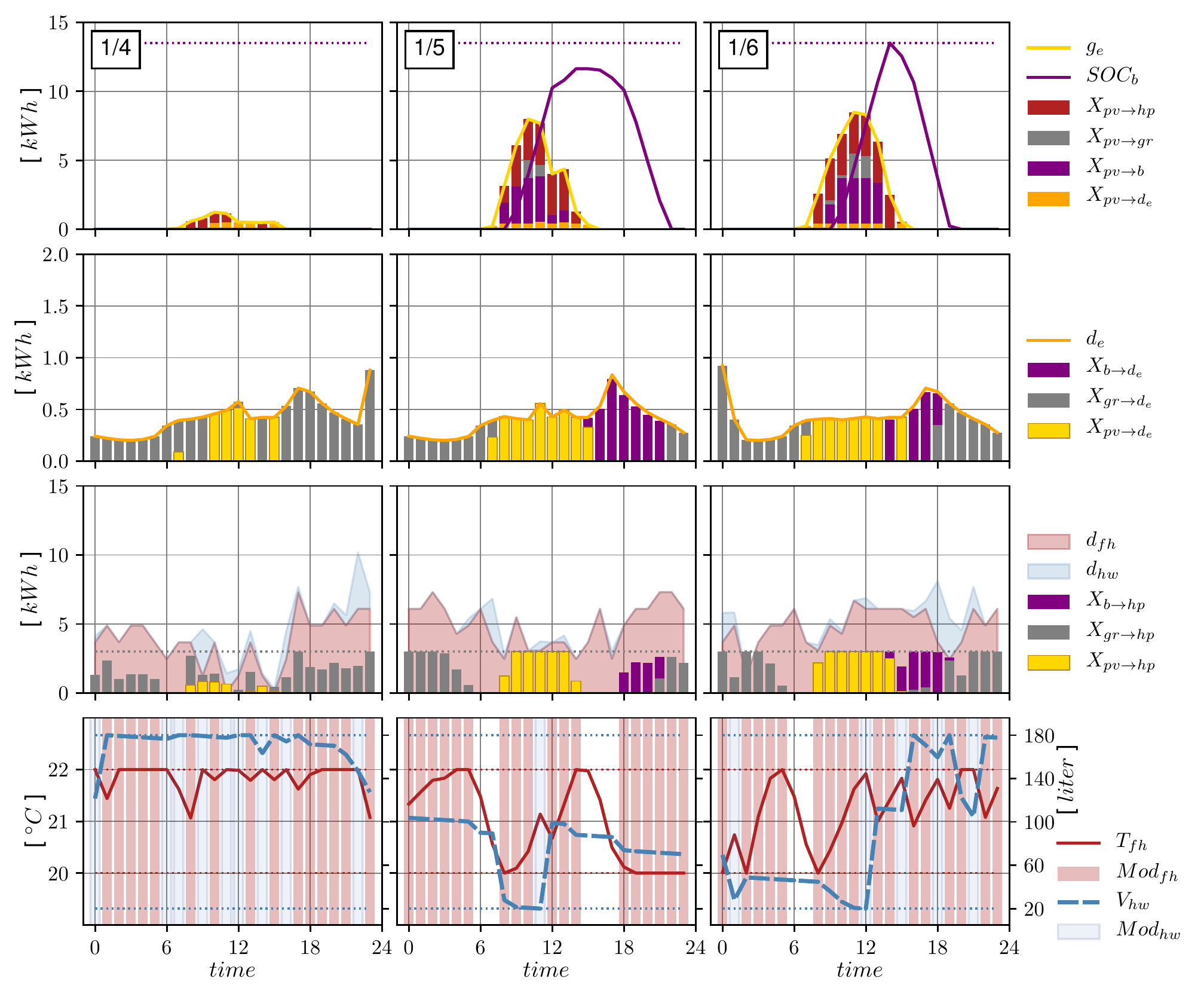}
		\caption{Flows from the PV system and demand fulfillment during winter days (base case).}
		\label{fig:Example_winter}
	\end{center}
\end{figure}

\subparagraph{Winter Edge Cases.}
The winter season is especially challenging because of the high heating demand and low PV generation. Figure~\ref{fig:Example_winter} shows the behavior of the system around the winter edge case (January~4--6). The PV production was exceptionally low on January~4.

The PV flows and battery SOCs (top row of Figure~\ref{fig:Example_winter}) are similar to those observed in Figure~\ref{fig:PV_Flows}. The dotted line denotes the maximum usable battery capacity. The second row of Figure~\ref{fig:Example_winter} plots the electricity demand and the sources fulfilling that demand (stacked bars). The third row denotes the thermal demands (areas under the curve), the sources of the heating energy (comprising stacked bars), and the modulation degree of the heat pump (bar heights). The dotted line indicates the maximum power of the heat pump. The $cop$ compensates the difference between the electrical and thermal energy. The bottom row illustrates the effect of the heat-pump operation on thermal SOCs. Further, the mode of the heat pump is indicated, i.e., whether it is off or running in the $fh$ or $hw$ modes. The left axis indicates the temperature of the $fh$ system (red line), whereas the right axis indicates the available volume in the $hw$ tank (dashed blue line). The dotted lines indicate the comfort ranges of both the systems.

On January~4~(column~1), the PV generation is insufficient to fulfill the electricity and thermal demands (rows~2 and~3, respectively). Hence, sourcing from the grid is necessary even during day (gray bars).

On January~5, both the thermal and electricity demands were fulfilled by PV generation. Subsequently, the battery was not fully charged (purple line in row 1) because the maximum nominal power of the inverter was reached 10--12\,h. The PV surplus was fed into the grid (gray bars in row~1).

On January~6, the demands were fulfilled and the battery was fully charged. At~9\,h, some grid feed-in was allowed. Again, the grid feed-in at~10--13\,h can be attributed to the power restrictions of the inverter. Charging was delayed as long as possible owing to the self-discharging losses of the battery. During the nighttime hours of all the days, the battery power was insufficient and energy was sourced from the grid to fulfill the demand.

\begin{figure}[ht]
	\begin{center}
		\includegraphics[width=12 cm]{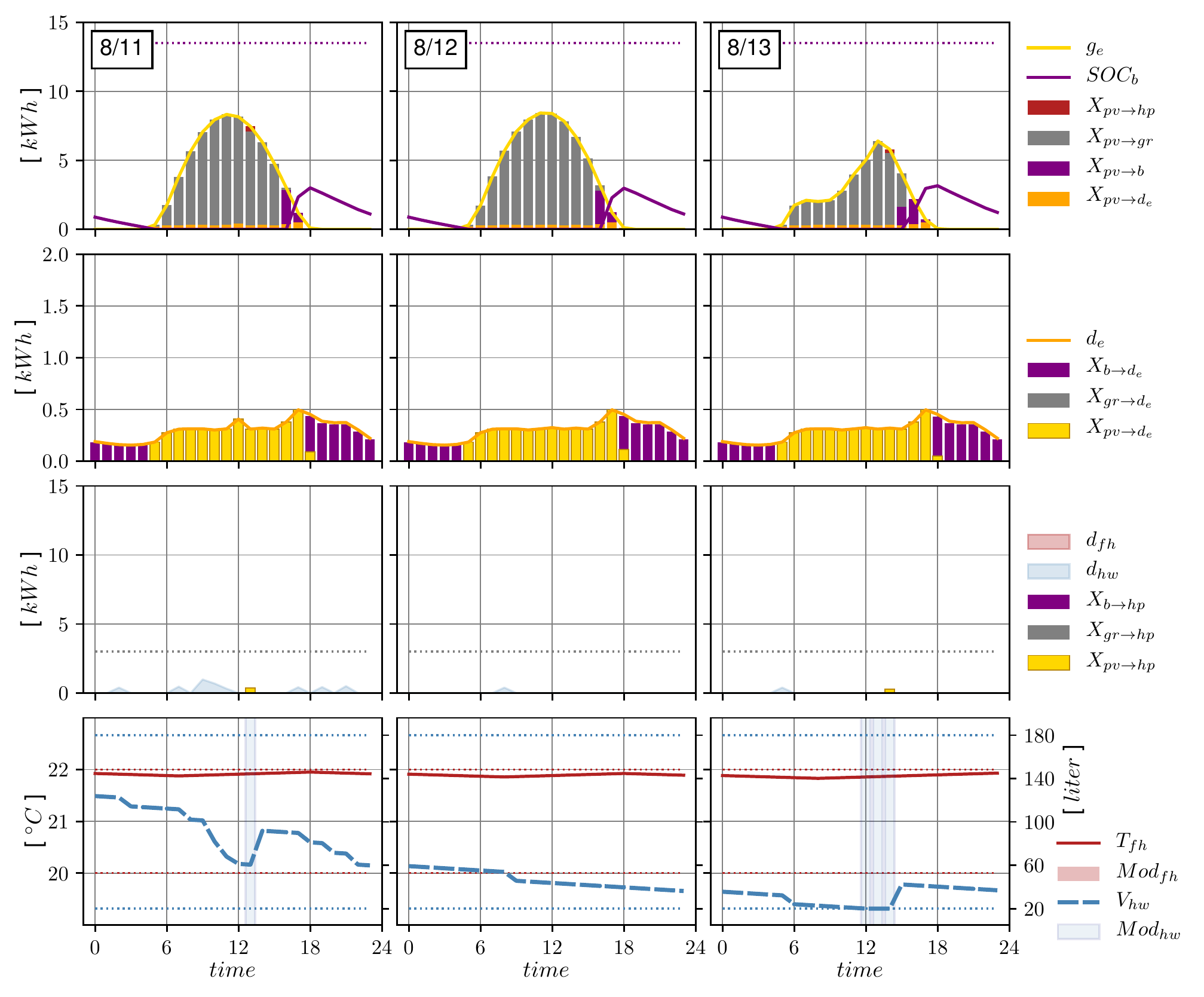}
		\caption{Flows from the PV system and demand fulfillment on three summer days (base case).}
		\label{fig:Example_summer}
	\end{center}
\end{figure}

\subparagraph{Summer Edge Cases.}
Figure~\ref{fig:Example_summer} presents the flows from the PV system and the electricity and thermal demands on three days in summer~(August 11--13) when the thermal demand was considerably low~(row~3). On all the three days, the electricity and thermal demands were fulfilled without sourcing from the grid. The battery was charged to its target SOC (approximately 25\% of the battery capacity; peaks of the purple line in row~1). The decisive factor in case of the target SOC was the electricity demand between the end of PV generation on the current day and the start of PV generation on the following day (purple bars in row~2). The PV surplus was fed into the grid.

As expected from Figure~\ref{fig:Heat_demand}, no floor heating was demanded in summer. The hot-water tank was heated solely based on PV generation. $cop_{hw}$ peaked when the outside temperature was the highest; accordingly, the tank was heated at~13\,h on August~11 and 14\,h on August~13. On the latter day, multiple very small charging instances could be observed immediately before the main charging instance. These were necessary to maintain the hot-water level (20\,liters) within the comfort ranges (row~4 of column~3). On August~12, when the hot-water demand was exceptionally low, charging can be avoided (row~4 of column~2). The benefits of the high $cop_{hw}$ on a warm day (August~11) must be counterbalanced against the depreciation losses of the water tank. On most days, charging once a day at the highest $cop_{hw}$ was optimal. The target SOC was obtained based on the forecasted demand until the next day.

Based on these observations and by assuming perfect information, both the rules of the target SOCs and the best time of the day for charging can be obtained. Considering the cost structure and the dissipation losses associated with the problem, the daily PV flows are fulfilled in the following order of priority:

\begin{enumerate}
	\item Electricity demand and target SOCs for heating
	\item Target SOC of the battery
	\item Feed-in to the grid
\end{enumerate}

The PV flows can be inferred from the target SOCs. The SOCs and charging times can be easily derived during summer. Obtaining the target SOCs during winter is more complicated because the solution space is constrained by the low PV generation and high heating demand. The following subsections present an in-depth analysis of the target SOCs.

\subsection{Target States of Charge} \label{sec:Target_SoC}
\begin{figure}[ht]
	\begin{center}
		\includegraphics[width=13.5 cm]{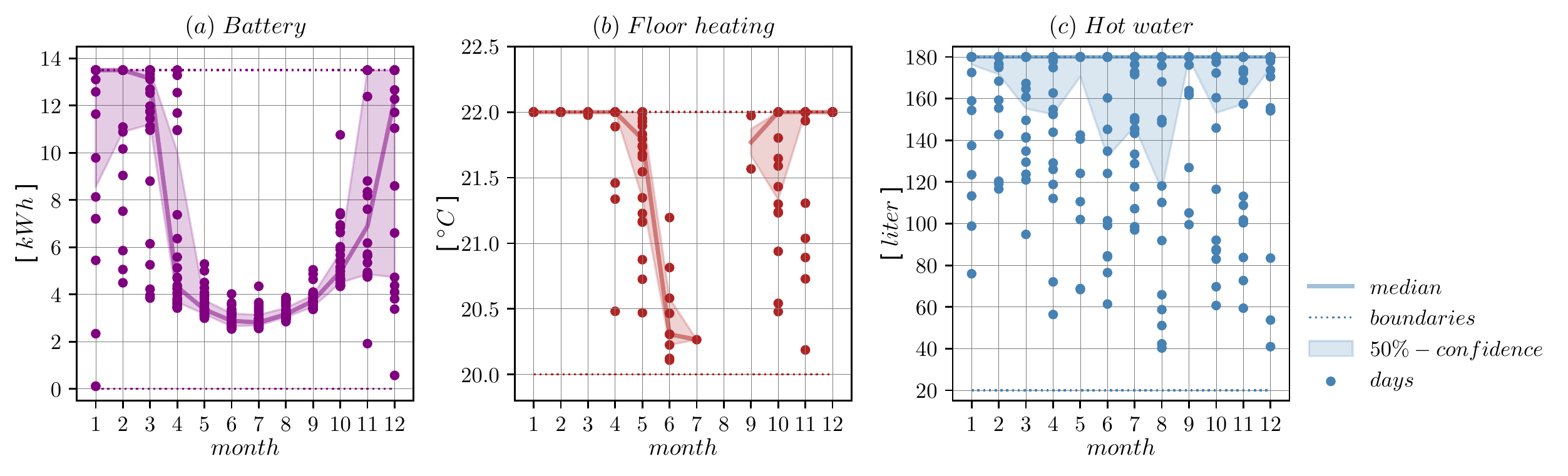}
		\caption{Target SOCs during each month (base case).}
		\label{fig:TSoCs_B+_GR+}
	\end{center}
\end{figure}

The target SOC is defined as the maximum value to which the storage should be charged on a given day. In Figure~\ref{fig:TSoCs_B+_GR+}, these daily values are illustrated for the perfect-information run (base case). An SOC is considered as a target SOC if charging was conducted in the previous time period because the change of state can only occur during the next time period. Therefore, because the floor was not heated in summer, the floor heating graph (Figure~\ref{fig:TSoCs_B+_GR+}~(b)) exhibited a low trend and less observations during the summer period. Apart from the daily target SOCs (points), Figure~\ref{fig:TSoCs_B+_GR+} plots the monthly medians (lines), 50\% confidence intervals (areas), and capacities or comfort bounds (dotted lines).

The maximum possible target SOC of the battery (Figure~\ref{fig:TSoCs_B+_GR+}~(a)) was defined based on the maximum usable capacity of the battery. Theoretically, the thermal target SOCs are restricted only by physical bounds and the penalties imposed on comfort-range violations. Charging is usually scheduled at the times during which peak $cop$ could be observed. 

\subparagraph{Battery Target SOCs.} 
The target SOCs of the battery exhibited a distinct U-shape (Figure~\ref{fig:TSoCs_B+_GR+}~(a)). The median values were close to the capacity limit during winter and decreased to become approximately a quarter of the capacity limit in summer. As noted before, this behavior is related to the battery characteristics. To avoid dissipation losses, charging should be conducted in accordance with the immediate upcoming demand (until the next PV generation), and the PV surplus should be fed into the grid. In winter, the target SOCs were raised accordingly when the floor-heating demand was high.

Observations lower than median values can be attributed to our definition of the target SOC. The reported values were obtained using the optimal solution and were subjected to modeling constraints. For instance, the battery charge capacity was sometimes limited by the PV generation or inverter capacity (column~2 of Figure~\ref{fig:Example_winter}). In these cases, the observed SOCs were lower than the unconstrained target SOCs. Because these instances mainly occurred in winter, the downward volatility was higher in winter compared with those during the remaining seasons.

Observations greater than the median values, which were especially common during the seasonal transitions in April and November, can be explained by the diverse floor-heating demands during these months. At the beginning of April, the battery is charged to a high target SOC to fulfill the high heating demand. By the end of April, the heating demands have reduced, resulting in lower target SOCs.

These results have interesting implications with respect to the design of rule-based control approaches. We suggest dynamic control of the target SOC values instead of setting a constant target SOC for the battery (as in common practice). These values can be defined based on the observed daily target SOC or the monthly or seasonal mean SOCs. The unconstrained target SOCs could be reasonably set to their daily maxima, i.e., their full capacity, to cope with the increased volatility observed during November and April. Dynamic target SOCs can potentially reduce the dissipation losses and increase the net profits when compared with the systems having constant values without jeopardizing the user comfort. Balancing the trade-off between the cost-saving potential and the information requirements of the alternating temporal aggregations should be attempted in future.

The aforementioned target SOCs are applicable only to charging from PVs and are inapplicable to charging from the grid. As explained earlier, when the electricity prices are constant, charging the battery from the grid is economically nonviable because of dissipation losses.

\subparagraph{Floor-heating Target SOCs.} 
In Figure~\ref{fig:TSoCs_B+_GR+}~(b), the floor-heating target SOCs are concentrated around the upper bound of the comfort range in winter, whereas they are considerably lower in summer when little or no heating is required. During the transition months, especially in May and October, volatility is introduced because of the fluctuating nature of the floor-heating demands. Similar to the battery and hot-water behaviors in Figure~\ref{fig:Example_summer}, the floor is heated by PV generation when it suffices. On high-demand days, the thermal storage is charged to its capacity limit.

From October to mid-November, the storage was charged (not necessarily to its upper limit) almost daily, preferably at the time of peak $cop_{fh}$. Smaller charging processes immediately before the main charging process may be necessary to satisfy the comfort constraint.

From mid-November to the winter season, the floor was typically heated to the upper-comfort bound and the PVs did not fully cover the heating demand. Therefore, additional power must be sourced from the grid to maintain the comfort constraint. The floor-heating control, especially in winter, aims to prevent comfort violations while utilizing the PV generation and battery. Accordingly, on most days, the SOC will be at the lower-comfort bound immediately before PV generation. When the PV generation starts, it can be fully utilized (see columns~2 and~3 of Figure~\ref{fig:Example_winter}). Based on these observations, we can derive the floor-heating target SOCs. 

\subparagraph{Hot-water Target SOCs.} 
Throughout the year, the hot-water demand is more stable than the floor-heating demands (see Figure~\ref{fig:Heat_demand}) because it is less influenced by the seasonal patterns. Therefore, the hot-water charging behavior is more evenly distributed during the year (Figure~\ref{fig:TSoCs_B+_GR+}~(c)).

As indicated in Figure~\ref{fig:Example_summer}, hot-water charging depends on the immediate upcoming demand. In summer, the demands on multiple days are sometimes aggregated to exploit the high $cop_{hw}$ on a hot day. In such cases, the target SOC is high (see row~4 in column~1 of Figure~\ref{fig:Example_summer}). The SOC is lower when the target demand is a single day’s usage (row~4 in column~3 of Figure~\ref{fig:Example_summer}). When the demanded floor heating is high, both the heat-pump modes compete for the maximum $cop$-hour because only one mode can operate at any one time. During the transition seasons, both the heat-pump modes attempt to utilize the limited PV generation. This effect slightly increases the target values. Especially when the $cop$ value reduces in winter, multiple heating periods are necessary to achieve a certain target SOC.\\

The observed target SOCs of the storage systems were observed within the bounds of capacity or thermal comfort. Within these ranges, the storage systems were charged under the constraints of PV generation and (in case of the battery) the inverter capacity. The target SOCs generally covered the demand up to the next PV generation. In winter, PV generation was insufficient, and the target SOCs were constrained to suboptimal values. Consequently, the target SOCs were widely diverse and differed on both daily and seasonal time scales. Regardless, they can be derived from the general pattern, and their target values should be set accordingly.

Given the important role of the demand values until the next PV generation, we analyze a rolling horizon approach with different prediction horizons in the following subsection.

\subsection{Prediction and Control Horizon} \label{sec:Rolling Horizon}
Until now, we have assumed perfect foresight over the whole time horizon (one year = 8760\,h). Perfect foresight means that the supply and demand data are known to the model. In a single optimization run, the solution time was almost 50\,min (using an Intel Core i7-8550U with an 1.8\,GHz processor, 4~cores, and 16~GB~RAM, allowing a mixed-integer programming gap of 0.5\%).

\begin{figure}[ht]
	\begin{center}
		\includegraphics[width=12 cm]{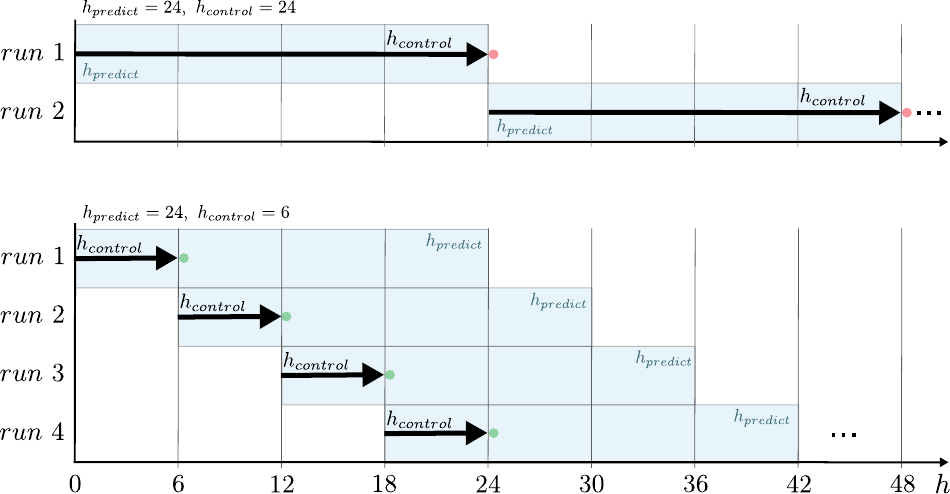}
		\caption{Rolling horizon approach showing the passed SOCs.}
		\label{fig:RollingHorizon_Graph}
	\end{center}	
\end{figure}

Most previous study assumed a prediction horizon of 24\,h \cite{Pean2019}. This modeling choice is justified by the data availability or the argument that long horizons do not significantly improve the results. The horizon is commonly set from midnight to midnight~\cite{Salpakari2016,Dengiz2019}. In the previous sections, we showed that the optimal charging amount is considerably dependent on the upcoming demand until the next PV generation. In the following subsection, we analyze the effect of restricted (perfect) foresight. We will evaluate the parameters $h_{predict}$ and $h_{control}$ associated with a rolling horizon approach (see Figure~\ref{fig:RollingHorizon_Graph}).

The prediction horizon $h_{predict}$ can be used to define the number of upcoming periods known to the model. The minimal prediction case is an online system in which the future information is ignored and fixed rules or schedules are applied to some current state, for example, based on the outside temperature. Although increasing the prediction horizon improves the model results (assuming perfect information), more data and a more advanced implementation are required. To analyze the value-of-information effects, we applied a rolling horizon approach with different $h_{predict}$ times~(24--96\,h). By fixing the control-horizon increment at 24\,h, we planned the next $h_{predict}$, fixed the first 24\,h, and set the resulting SOCs at (24+1)\,h as the initial states of the subsequent optimization run. This process for $h_{predict}$=24\,h is illustrated in the top part of Figure~\ref{fig:RollingHorizon_Graph}. To achieve comparable results, the reported evaluation horizon was standardized from January~1 to December~27~(8664\,h) because $h_{predict}$=4\,days can be observed only during this period.

The control horizon $h_{control}$ defines the number of time periods over which planned decisions are executed before beginning a new optimization run. The prediction and control horizons are often uniformly set to 24\,h (as in Figure~\ref{fig:RollingHorizon_Graph}~(top)). However, this setting may result in unintended end-of-horizon effects. When the whole prediction horizon is fixed at 24\,h, the positive SOCs at the end of the horizon add no value to the model. For example, the battery will always be empty at midnight; therefore, it cannot cover the morning demand before the first PV generation. In the bottom part of Figure~\ref{fig:RollingHorizon_Graph}, $h_{control}$ is reduced from 24\,h to 6\,h while maintaining $h_{predict}$ at 24\,h. In this setting, the end-of-horizon effect is mitigated when compared with that observed in case of an equal-length prediction and control horizon. To analyze the end-of-horizon effects for different values of $h_{predict}$, we varied $h_{control}$ as 24,~12, and~6\,h in the rolling horizon approach.

\begin{table}[ht]
	\centering
	\caption{Annual KPIs in the rolling horizon approach with varying prediction horizon (base case).}\label{tab:KPIs_rollingHorizon}
	\begin{tabular}{ l c| c c c c }
		\hline
		$h_{predict}$                         & 8664 & 24   & 36   & 48   & 96   \\ \hline
		$h_{control}$                         & 8664 & \multicolumn{4}{|c}{24}   \\ \hline
		Overall energy consumption [kWh/year] & 7357 & 7440 & 7374 & 7371 & 7368 \\
		Self-sufficiency rate [\%]            & 80   & 64   & 79   & 79   & 79   \\
		Overall profit [EUR/year]             & 595  & 360  & 583  & 586  & 592  \\ \hline
		Sum of violations                     & 79   & 2022 & 86   & 82   & 82   \\ \hline
		Runtime [min]                         & 40   & 0.13 & 0.14 & 0.19 & 0.48 \\ \hline
	\end{tabular} 
\end{table}

\subparagraph{Prediction horizon.} 
The results are compared in Table~\ref{tab:KPIs_rollingHorizon}. Compared to the full-horizon model (8664\,h), the rolling horizon approach slightly increased the energy consumption (by 1\% at 24\,h and 0.1\% at 96\,h), reduced the self-sufficiency rate (by 20\% at 24\,h and 1\% at 96\,h), lowered the profits (by 40\% at 24\,h and 0.5\% at 96\,h), and increased the proportion of comfort violations (by 2460\% at 24\,h and 4\% at 96\,h). All the KPIs were improved by extending the prediction horizon; however, the improvement rates decreased and the runtimes increased. Nevertheless, the runtime (even for $h_{predict}$=4\,days) was significantly lower in the rolling horizon model than that in the full-horizon model. Based on the results in Table~\ref{tab:KPIs_rollingHorizon}, the 24-h prediction horizon conferred no benefits, especially with respect to comfort violations. The large gap between the KPIs of the 24- and 36-h cases may denote the essential role of knowing the demand until the next PV generation. However, the end-of-horizon effects may also be the cause, as shown in the next subsection.

As long prediction horizons require more data, the trade-off between the forecasting costs and data quality should be considered along with the overall improvements, and the parameters should be appropriately set.

\begin{table}[ht]
	\centering
	\caption{Annual KPIs in the rolling horizon approach with varying control horizon (base case).}\label{tab:KPIs_rollingHorizon_fix}
	\begin{tabular}{ l c | c c | c c | c c| c c}
		\hline
		$h_{predict}$     & 8664 & \multicolumn{2}{|c|}{24} & \multicolumn{2}{|c|}{36} & \multicolumn{2}{|c|}{48} & \multicolumn{2}{|c}{96} \\ \hline
		$h_{control}$     & 8664 & 12   & 6                 & 12   & 6                 & 12   & 6                 & 12   & 6                \\ \hline
		Consumption       & 7357 & 7377 & 7375              & 7372 & 7370              & 7369 & 7368              & 7368 & 7367             \\
		Self-sufficiency  & 80   & 75   & 78                & 79   & 79                & 79   & 79                & 80   & 80               \\
		Profit            & 595  & 528  & 571               & 585  & 585               & 588  & 590               & 591  & 591              \\ \hline
		Violations        & 79   & 86   & 86                & 82   & 82                & 82   & 82                & 82   & 82               \\
		Runtime           & 40   & 0.24 & 0.44              & 0.29 & 0.58              & 0.38 & 0.77              & 0.98 & 2.01             \\ \hline
	\end{tabular}    
\end{table}

\subparagraph{Control horizon.} 
The second important parameter of the rolling horizon scheme is $h_{control}$. In the previous analyses, the first 24\,h was fixed for all the lengths of the prediction horizon. In this scheme, setting $h_{predict}$=24\,h was disadvantageous, and longer horizons yielded superior results. In case of the 24-h prediction horizon, the final hour lacked any foresight (see Figure~\ref{fig:RollingHorizon_Graph}~(top)). Owing to the end-of-horizon effects, the storage systems were always empty at the end of each optimization run. Conversely, the 36-h look-ahead captured not only the final hour of the horizon (24\,h) but also the subsequent 12\,h foresight, mitigating the negative effects.

The end-of-horizon effects (even those of the 24-h prediction horizon) were mitigated by adapting $h_{control}$. The KPIs of the 24-h prediction case were significantly improved when the length of the control horizon was reduced. In the 6-h control horizon, the results almost reached those of the 36-h prediction case because the final hour (6\,h) considers the subsequent 16\,h foresight (see Figure~\ref{fig:RollingHorizon_Graph}~(bottom)). However, with the increasing lengths of the prediction horizons, the improvement rates decreased, and in the 96-h prediction case no significant improvements were achieved. Increasing the length of the prediction horizon also extended the runtimes. The severity of the end-of-horizon effects and the widely used unfavorable parameter settings necessitate further investigations.

\section{Conclusions} \label{sec:Conclusions}
In this study, we identified the structural properties of the optimal operating policy of the MPC algorithm for a SHEMS in a single residential building. The energy system contains a modulating air-to-water heat pump (maximum power:~3\,kW), a PV system (capacity:~10\,kWp), a battery (nominal capacity:~14\,kWh), and thermal storage systems for floor heating and hot-water supply. We allow grid feed-in and sourcing by fixing the feed-in tariffs and retail prices at their current values in Germany.

Based on our numerical analysis of four technological configurations, we observed that the battery was essential to improve the self-consumption and self-sufficiency rates of the system. Without a battery, solely exploiting the flexibility of the thermal storage system, the self-consumption and self-sufficiency rates were only 24\% and 52\%, respectively. After the installation of a battery, these rates increased to 37\% and 79\%, respectively. In addition, we observed that without grid feed-in PV curtailment was unavoidable under the representative demand pattern and PV battery system. The grid feed-in improved the processing efficiency of the PV surplus, decreasing the overall energy consumption without deteriorating the self-sufficiency levels and causing comfort violations. When feed-in tariffs are absent or decreasing or when the grid is congested, additional revenue streams by peer-to-peer trading could be considered. The control, regulation, and fairness of this alternative should be further investigated.

A commonly used objective (maximizing the self-consumption) was observed to be economically nonviable considering the feed-in tariffs in Germany. The maximization of self-consumption reduced the efficiency of processing PV generation. The slight increase in self-consumption after maximization (from 37\% to 39\%) was offset by increased losses and energy consumption. Moreover, because the feed-in was reduced, the net profits were 9\% lower than those of cost minimization. The maximization of self-sufficiency improved the performance compared to the self-consumption case and reduced the solution time from approximately 50~to~10\,min; however, it marginally reduced the profits (by approximately 2\%) below those of cost minimization. The energy consumption decreased compared to the self-consumption case because storage losses were considered. Only applying the cost-minimizing objective balanced the trade-off between the efficiency loss and the potential revenue from grid feed-in appropriately.

By analyzing the optimal flows and cost structures, we determined the dominant order in which the available PV generation on any given day is distributed within the integrated system. First, the electricity demand was fulfilled, and the heating SOCs were satisfied. The target SOCs define the levels to which the storage units should be charged. Next, the battery target SOC was fulfilled; finally, the PV surplus was fed into the grid. We quantified the target SOCs and determined the best time for charging the storage devices. The target SOCs were determined based on the immediately upcoming demand until the next PV generation (on the next day). Because the floor-heating demand was seasonally variable, the battery and floor-heating target SOCs widely varied among the summer, winter, and transitional periods; in contrast, the hot-water target SOCs remained constant throughout the year. The times of charging during PV hours were especially distinct for heating, charging is mainly determined based on the maximum coefficient of performance. Thus, charging is mainly conducted at times of peak outdoor temperatures. Under the derived rules, the target SOCs can be quantified on daily, monthly, or seasonal time scales, improving the commonly fixed rule-based approaches. The target SOC performances on different time scales should be evaluated via a simulation study based on data obtained from different years.

By applying a rolling horizon approach, we reduced the solution time from approximately 50\,min to less than 1\,min (obtaining the solution for one year with a time resolution of 1\,h). We evaluated the value of information for different prediction horizons and the end-of-horizon effects for different control horizons. The commonly used prediction horizon (24\,h with a control horizon of also 24\,h) resulted in inefficient system behavior. We showed that the end-of-horizon effects are easily mitigated with no further information requirements by reducing the control horizon.

The following limitations and potential expansions are underway: adjusting the algorithms to handle the uncertainties in demand and generation~\cite{Dengiz2019a}, investigating more complex heat pump representations~\cite{Fischer2017}, developing self-learning algorithms that incorporate stochastic behavior and nonlinearities and that adapt to changing user behaviors~\cite{Yu2020,VazquezCanteli2019}, and creating multiagent systems in which energy communities can trade in a peer-to-peer market~\cite{Lueth2018, Neves2020,Nguyen2018}.

By applying the proposed formulations and insights, other researchers can reduce the action space associated with complex stochastic algorithms such as those of the self-learning systems. In these model formulations, reducing the decision space can mitigate the curse of dimensionality in both continuous-state and continuous-action spaces as with modulating heat pumps and continuous storage-system charging.

\bibliographystyle{plainnat}
\bibliography{ms}








\appendix

\clearpage
\section{Variables and Model Parameters.} \label{sec:Parameters}

\begin{table}[ht]
	\centering
	\caption{Decision variables of the mathematical model}\label{tab:variables}
	\begin{tabular}{ p{7cm} p{6cm} }
		\hline
		\multicolumn{2}{l}{\textbf{Flow variables}}                                                                                                \\ \hline
		Demand fulfillment [kWh]                         & $X_{pv\rightarrow d_e}(h), \ X_{b\rightarrow d_e}(h), \ X_{gr\rightarrow d_e}(h)\geq 0$ \\
		Heat pump inflows [kWh]                          & $X_{pv\rightarrow hp}(h), \ X_{gr\rightarrow hp}(h),\  X_{b\rightarrow hp}(h)\geq 0$    \\
		Heat pump modulation [kWh]                       & $X_{hp\rightarrow fh}(h), \ X_{hp\rightarrow hw}(h) \geq 0$                             \\
		Other PV output [kWh]                               & $X_{pv\rightarrow b}(h), \ X_{pv\rightarrow gr}(h)\geq 0$                               \\ \hline
		\multicolumn{2}{l}{\textbf{States of charge}}                                                                                              \\ \hline
		State of charge of the battery [kWh]                 & $SOC_b(h) \geq 0$                                                                       \\
		Temperature of the $fh$ system [$^\circ$C]           & $T_{fh}(h) \geq 0$                                                                      \\
		Volume of available $hw$ [l]                     & $V_{hw}(h) \geq 0$                                                                      \\ \hline
		\multicolumn{2}{l}{\textbf{Heat pump variables}}                                                                                           \\ \hline
		Modulation degree of $fh$ [\%]                   & $Mod_{fh}(h) \geq 0$                                                                    \\
		Modulation degree of $hw$ [\%]                   & $Mod_{hw}(h) \geq 0$                                                                    \\
		Heat pump mode (binary)                            & $HP^{switch}(h) \in [0,1]$                                                              \\
		Comfort violations $fh$ (pos./neg.) [$^\circ$ C] & $T_{fh}^+(h), \ T_{fh}^-(h) \geq 0$                                                     \\
		Temperature loss per time period (+/-)           & $Loss_{fh}^{+/-}(h) \geq 0$                                                             \\
		Comfort violations $hw$ (pos./neg.) [l]          & $V_{hw}^+(h), \ V_{hw}^-(h) \geq 0$                                                     \\
		Hot outside temperature (binary)                   & $Hot(h) \in [0,1]$                                                                      \\ \hline
		\multicolumn{2}{l}{\textbf{Additional variables for alternative objectives}}                                                                 \\ \hline
		Heat pump mode (binary)                            & $B^{switch}(h) \in [0,1]$                                                               \\ \hline
	\end{tabular} 
\end{table}

\begin{table}[ht]
	\centering
	\caption{Specifications of the photovoltaic system (based on \cite{Speichermonitoring2018})}\label{tab:pv_specs}
	\begin{tabular}[ht]{ p{7cm} p{3cm} p{3cm}}
		\hline
		\textbf{Definition}             & \textbf{Parameter} & \textbf{Value} \\ \hline
		Capacity of the photovoltaic system & $pv^{max}$         & 10\,kWp         \\
		System loss                     &                    & 0              \\
		Year                            &                    & 2015           \\
		Location                        &                    & Chicago        \\
		Tilt                            &                    & 35$^{\circ}$  \\
		Azimuth                         &                    & 180$^{\circ}$ \\ \hline
	\end{tabular} 
\end{table}

\begin{table}[ht]
	\centering
	\caption{Specifications of the heat pump system (based on \cite{Dengiz2019})}\label{tab:HP_specs}
	\begin{tabular}[ht]{ p{7cm} p{3cm} p{3cm}}
		\hline
		\textbf{Definition}                     & \textbf{Parameter} & \textbf{Value}                    \\ \hline
		Maximal power of the heat pump          & $hp^{max}$         & 3\,kW                             \\ \hline
		\multicolumn{3}{l}{\textbf{Floor heating}}                                                       \\ \hline
		Capacity of the floor heating system      & $v_{fh}$           & 10\,$m^3$                         \\
		Density of concrete                     & $p_{concr}$        & 2400\,$\frac{kg}{m^3}$            \\
		Heat capacity of concrete               & $c_{concr}$        & 1\,$\frac{kJ}{kg*^{\circ} C}$     \\
		Supply temperature                      & $t_{fh}^{supply}$  & 30$^{\circ}$C                   \\
		Lower bound of the comfort range        & $t_{fh}^{min}$     & 20$^{\circ}$C                   \\
		Upper bound of the comfort range        & $t_{fh}^{max}$     & 22$^{\circ}$C                   \\
		Temperature loss per time period (+/-)  & $loss_{fh}$        & 0.045\,kW                         \\
		Coefficient of performance              & $cop_{fh}(h)$      & $[8760\times 1]$                  \\
		Maximum temperature difference inside/outside & $big$              & 60$^{\circ}$C                   \\ \hline
		\multicolumn{3}{l}{\textbf{Hot water}}                                                           \\ \hline
		Supply temperature of the hot water            & $t_{hw}^{supply}$  & 45$^{\circ}$C                   \\
		Density of water                        & $p_{water}$        & 997\,$\frac{kg}{m^3}$             \\
		Heat capacity of water                  & $c_{water}$        & 4.184\,$\frac{kJ}{kg*^{\circ} C}$ \\
		Lower bound of the comfort range        & $v_{hw}^{min}$     & 20\,l                             \\
		Upper bound of the comfort range        & $v_{hw}^{max}$     & 180\,l                            \\
		Volume loss per time period             & $loss_{hw}$        & 0.035\,kW                         \\
		Coefficient of performance              & $cop_{hw}(h)$      & $[8760\times 1]$                  \\ \hline
	\end{tabular} 
\end{table}

\begin{table}[ht]
	\centering
	\caption{Specifications of the battery system (Tesla Powerwall 2 \cite{Tesla2018})}\label{tab:battery_specs}
	\begin{tabular}[ht]{ p{7cm} p{3cm} p{3cm}}
		\hline
		\textbf{Definition}                       & \textbf{Parameter} & \textbf{Value} \\ \hline
		Minimum usable capacity                       & $soc_b^{min}$      & 0\,kWh          \\
		Maximum usable capacity (nominal=14\,kWh)    & $soc_b^{max}$      & 13.5\,kWh       \\
		One-way efficiency (discharging/charging)        & $\eta_b$           & 95\%           \\
		Nominal power of inverter (discharging/charging) & $b^{max}$          & 3.3\,kW         \\
		Maximal dis-/charging per hour          & $b_{rate}^{max}$   & 3.3\,kWh        \\
		Loss of battery capacity per time period       & $loss_b$           & 0.003\%       \\ \hline
	\end{tabular} 
\end{table}

\begin{table}[ht]
	\centering
	\caption{Specifications of exogenous data} \label{tab:data}
	\begin{tabular}[ht]{ p{7cm} p{3cm} p{3cm}}
		\hline
		\textbf{Definition}              & \textbf{Parameter} & \textbf{Size/Source}                   \\ \hline
		Electricity demand [kWh]         & $d_e(h)$           & $[8760\times1]$ \cite{Christensen2006} \\
		PV generation [kWh]              & $g_e(h)$           & $[8760\times1]$ \cite{Pfenninger2016}  \\
		Floor heating demand [kWh]       & $d_{fh}(h)$        & $[8760\times1]$ \cite{Christensen2006} \\
		Hot water demand [kWh]           & $d_{hw}(h)$        & $[8760\times1]$ \cite{Christensen2006} \\
		Outside temperature [$^\circ C$] & $t_{outside}(h)$   & $[8760\times1]$ \cite{Rienecker2011}   \\ \hline
	\end{tabular} 
\end{table}

\begin{table}[ht]
	\centering
	\caption{Specifications of tariffs and cost factor}\label{tab:prices}
	\begin{tabular}[ht]{p{7cm} p{2cm} p{4cm}}
		\hline
		\textbf{Definition}                                 & \textbf{Parameter} & \textbf{Value/Source}                                        \\ \hline
		Retail price (from grid)                            & $p_{buy}$          & 0.30\,EUR/kWh  \cite{GreenpeaceEnergy2020, NaturstromAG2020} \\
		Grid feed-in tariff (to grid)                       & $p_{sell}$         & 0.10\,EUR/kWh  \cite{Bundesnetzagentur2020}                  \\
		Cost per unit of comfort violation ($^\circ C$/ l) & $costfactor$       & 1\,EUR/Unit  \cite{Yu2020, Masy2015}                         \\ \hline
	\end{tabular} 
\end{table}

\clearpage
\section{Mathematical model.} \label{sec:Model}

\textbf{Objective function:}
\begin{align}
max &\sum_{h \in H} \left( p_{sell} * X_{pv\rightarrow gr}(h) - p_{buy} * \left( X_{gr\rightarrow d_e}(h) + X_{gr\rightarrow hp}(h) \right) - C_{violation}(h) \right) 		\label{eq:Objective}\\
&where \quad C_{violation}(h)= cost~factor* \left( T_{fh}^+(h) + T_{fh}^-(h) + V_{hw}^+(h) + V_{hw}^-(h) \right) 
\\ \nonumber
\end{align}

\textbf{s.t.} 

\textbf{Flow constraints:}
\begin{align}
X_{pv\rightarrow d_e}(h) + X_{b\rightarrow d_e}(h) + X_{gr\rightarrow d_e}(h) = d_e(h),\ \forall h& 		\label{eq:DE_fulfillment} \\
X_{pv\rightarrow d_e}(h) + X_{pv\rightarrow b}(h) + X_{pv\rightarrow gr}(h) + X_{pv\rightarrow hp}(h)   = g_e(h),\ \forall h&  			\label{eq:GE_restriction}
\\ \nonumber
\end{align}

\textbf{Battery constraints:}
\begin{align}
SOC_b(h+1) = \left(1-loss_b\right)*SOC_b(h) +  \eta_b*X_{pv\rightarrow b}(h) -\left(X_{b\rightarrow d_e}(h) + X_{b\rightarrow hp}(h)\right)/\eta_b,\ \forall h&  																													\label{eq:B_SOC} \\
soc_b^{min} \leq SOC_b(h) \leq soc_b^{max},\ \forall h& 																			\label{eq:SOC_B_min_max} \\
X_{b\rightarrow d_e}(h), \ X_{pv\rightarrow b}(h), \ X_{b\rightarrow hp}(h) \leq b_{rate}^{max},\ \forall h&									\label{eq:B_SOC_charge_discharge}
\\ \nonumber
\end{align}

\textbf{Heat pump constraints:}
\begin{align}
X_{hp\rightarrow fh}(h) + X_{hp\rightarrow hw}(h) &= X_{pv\rightarrow hp}(h) + X_{gr\rightarrow hp}(h) + X_{b\rightarrow hp}(h),\ \forall h  	\label{eq:HP_balance} \\ 
\nonumber \\
X_{hp\rightarrow fh}(h) &= Mod_{fh}(h) * hp^{max},\ \forall h																			\label{eq:HP_modulation} \\
X_{hp\rightarrow hw}(h) &= Mod_{hw}(h) * hp^{max},\ 	\forall h \nonumber \\
\nonumber \\
Mod_{fh}(h) &\leq HP^{switch}(h),\	\forall h 																							\label{eq:HP_mode} \\
Mod_{hw}(h) &\leq 1 - HP^{switch}(h),\ \forall h \nonumber
\\ \nonumber
\end{align}

\textbf{Floor-heating constraints:}
\begin{align}
T_{fh}(h+1) = T_{fh}(h) + conv_{fh}*\left( cop_{fh}(h) * X_{hp\rightarrow fh}(h) -d_{fh}(h) - Loss_{fh}^{+/-}  \right),& \forall h& 			\label{eq:FH_next}
\end{align}

\begin{align}
where \quad conv_{fh} &= \dfrac{60 * 60}{p_{concr}*v_{fh}*c_{concr}}				\label{eq:FH_conv} \\
Loss_{fh}^{+/-} &= \left(1-Hot(h)\right)*loss_{fh} - Hot(h)*loss_{fh},\ \forall h  			\label{eq:FH_loss}
\end{align}

\begin{align}
T_{fh}(h) - \left(1 - Hot(h)\right)*big &\leq t_{outside}(h),\ \forall h  												\label{eq:FH_hot}\\
t_{outside}(h) - Hot(h)*big &\leq T_{fh}(h) ,\ \forall h \nonumber \\ 
\nonumber\\
T_{fh}(h) &\leq t_{fh}^{max} + T_{fh}^+(h),\ \forall h  												\label{eq:FH_min_max}\\
t_{fh}^{min} - T_{fh}^-(h) &\leq T_{fh}(h),\ \forall h \nonumber 
\\ \nonumber
\end{align}

\textbf{Hot-water constraints:}
\begin{align}
V_{hw}(h+1) = V_{hw}(h) + conv_{hw}* \left(cop_{hw}(h) * X_{hp\rightarrow hw}(h) - d_{hw}(h) - loss_{hw} \right),&\ \forall h& 			\label{eq:HW_next}\end{align}

\begin{align}
&where \quad conv_{hw} = \dfrac{60*60}{\left(p_{water}*t_{hw}^{supply}*c_{water}\right)/1000}						\label{eq:HW_conv}
\end{align}

\begin{align}
V_{hw}(h) &\leq v_{hw}^{max} + V_{hw}^+(h),\ \forall h  						\label{eq:HW_min_max}\\
v_{hw}^{min} - V_{hw}^-(h) &\leq V_{hw}(h),\ \forall h  \nonumber 
\\ \nonumber
\end{align}

\end{document}